\documentclass[11pt]{article}
\usepackage{geometry}                % See geometry.pdf to learn the layout options. There are lots.
\usepackage{graphicx}
\usepackage{amssymb}
\usepackage{amsmath} 				% Needed for align environment
\usepackage{epstopdf}

\usepackage{latexsym}
\usepackage{marvosym}
\usepackage{bm}	%This is for bold greek letter and such; command is \bm

\usepackage[mathcal]{euscript}
\usepackage[all]{xy}
\usepackage{indentfirst}

\numberwithin{equation}{section}

\usepackage{hyperref}
\hypersetup{
	colorlinks,
	citecolor=cyan,% filecolor=black,% 
	linkcolor=cyan,% urlcolor=black,% pdftex
	bookmarksopen=true,
	bookmarksopenlevel=\maxdimen,
	}

\usepackage{amsthm} %Enable theorem environments
\theoremstyle{definition}

\theoremstyle{plain}

\def\K3{\mathrm K3}

\def\double #1{#1{\hbox{\kern-2pt $#1$}}}
\def\pp{{\mathchoice
          {%displaystyle
              \kern 1pt%
              \raise 1pt
              \vbox{\hrule width5pt height0.4pt depth0pt
                    \kern -2pt
                    \hbox{\kern 2.3pt
                          \vrule width0.4pt height6pt depth0pt
                          }
                    \kern -2pt
                    \hrule width5pt height0.4pt depth0pt}%
                    \kern 1pt
           }
            {%textstyle
              \kern 1pt%
              \raise 1pt
              \vbox{\hrule width4.3pt height0.4pt depth0pt
                    \kern -1.8pt
                    \hbox{\kern 1.95pt
                          \vrule width0.4pt height5.4pt depth0pt
                          }
                    \kern -1.8pt
                    \hrule width4.3pt height0.4pt depth0pt}%
                    \kern 1pt
            }
            {%scriptstyle
              \kern 0.5pt%
              \raise 1pt
              \vbox{\hrule width4.0pt height0.3pt depth0pt
                    \kern -1.9pt  %[e]=0.15pt
                    \hbox{\kern 1.85pt
                          \vrule width0.3pt height5.7pt depth0pt
                          }
                    \kern -1.9pt
                    \hrule width4.0pt height0.3pt depth0pt}%
                    \kern 0.5pt
            }
            {%scriptscriptstyle
              \kern 0.5pt%
              \raise 1pt
              \vbox{\hrule width3.6pt height0.3pt depth0pt
                    \kern -1.5pt
                    \hbox{\kern 1.65pt
                          \vrule width0.3pt height4.5pt depth0pt
                          }
                    \kern -1.5pt
                    \hrule width3.6pt height0.3pt depth0pt}%
                    \kern 0.5pt%}
            }
        }}

\def\mm{{\mathchoice
                       {%displaystyle
                             \kern 1pt
               \raise 1pt    \vbox{\hrule width5pt height0.4pt depth0pt
                                  \kern 2pt
                                  \hrule width5pt height0.4pt depth0pt}
                             \kern 1pt}
                       {%textstyle
                            \kern 1pt
               \raise 1pt \vbox{\hrule width4.3pt height0.4pt depth0pt
                                  \kern 1.8pt
                                  \hrule width4.3pt height0.4pt depth0pt}
                             \kern 1pt}
                       {%scriptstyle
                            \kern 0.5pt
               \raise 1pt
                            \vbox{\hrule width4.0pt height0.3pt depth0pt
                                  \kern 1.9pt
                                  \hrule width4.0pt height0.3pt depth0pt}
                            \kern 1pt}
                       {%scriptscriptstyle
                           \kern 0.5pt
             \raise 1pt  \vbox{\hrule width3.6pt height0.3pt depth0pt
                                  \kern 1.5pt
                                  \hrule width3.6pt height0.3pt depth0pt}
                           \kern 0.5pt}
                       }}

\def\ad{{\kern0.5pt
                   \alpha \kern-5.05pt
\raise5.8pt\hbox{$\textstyle.$}\kern 0.5pt}}

\def\bd{{\kern0.5pt
                   \beta \kern-5.05pt \raise5.8pt\hbox{$\textstyle.$}\kern 0.5pt}}

\def\qd{{\kern0.5pt
                   q \kern-5.05pt \raise5.8pt\hbox{$\textstyle.$}\kern 0.5pt}}
\def\Dot#1{{\kern0.5pt
     {#1} \kern-5.05pt \raise5.8pt\hbox{$\textstyle.$}\kern 0.5pt}}

\font\ro=cmsy10                          % font with rope
\def\kcr{{\hbox{\ro \char'170}}}                % right-handed rope
\def\ktl{{\hbox{\ro \char'170}}}        % top end for left-handed rope
\def\ktr{{\hbox{\ro \char'170}}}        % " right
\def\kbl{{\hbox{\ro \char'170}}}        % " bottom left
\def\kbr{{\hbox{\ro \char'170}}}        % " right

\def\border{                                            % border
        \setlength{\unitlength}{1mm}
        \newcount\xco
        \newcount\yco
        \xco=-21
        \yco=12
        \begin{picture}(140,0)
        \put(\xco,\yco){$\ktl$}
        \advance\yco by-1
        {\loop
        \put(\xco,\yco){$\kcr$}
        \advance\yco by-2
        \ifnum\yco>-240
        \repeat
        \put(\xco,\yco){$\kbl$}}
        \xco=158
        \yco=12
        \put(\xco,\yco){$\ktr$}
        \advance\yco by-1
        {\loop
        \put(\xco,\yco){$\kcr$}
        \advance\yco by-2
        \ifnum\yco>-240
        \repeat
        \put(\xco,\yco){$\kbr$}}
        \put(-20,13){\tiny **University of Maryland * Center for String and
         Particle  Theory* Physics Department***University of Maryland *Center
        for String and Particle  Theory** }
        \put(-20,-241.5){\tiny **University of Maryland * Center for String and
         Particle  Theory* Physics Department***University of Maryland *Center
        for String and Particle  Theory** }
        \end{picture}
        \par\vskip-8mm}

\def\headpic{                                           % UM heading
        \indent
        \setlength{\unitlength}{.4mm}
        \thinlines
        \par
        \begin{picture}(29,16)
        \put(165,16){\line(1,0){4}}
        \put(170,16){\line(1,0){4}}
        \put(180,16){\line(1,0){4}}
        \put(175,0){\line(1,0){4}}
        \put(180,0){\line(1,0){4}}
        \put(185,0){\line(1,0){4}}
        \put(169,0){\line(0,1){16}}
        \put(170,0){\line(0,1){16}}
        \put(179,0){\line(0,1){16}}
        \put(180,0){\line(0,1){16}}
        \put(184,0){\line(0,1){16}}
        \put(185,0){\line(0,1){16}}
        \put(169,16){\oval(8,32)[bl]}
        \put(170,16){\oval(8,32)[br]}
        \put(179,0){\oval(8,32)[tl]}
        \put(185,0){\oval(8,32)[tr]}
        \end{picture}
        \par\vskip-6.5mm
        \thicklines}
\def\endtitle{\end{quotation}\newpage}                  % end title page

\def\title{Superforms in  Simple Six-dimensional Superspace}
\def\author{Cesar Arias$,\hspace{-5pt}{}^{\small \mbox\Scorpio}$ William D.~Linch {\sc iii}$,\hspace{-5pt}{}^{\small \mbox\Pisces}$ and Alexander K. Ridgway$~\hspace{-5pt}{}^{\small \mbox\Virgo}$}

\begin{document}

\thispagestyle{empty}

\border\headpic {\hbox to\hsize{\today \hfill
{UMDEPP-014-003}}}
\par \noindent
~%{ \hfill{arXiv:xxxx.xxxx [hep-th]}}
\par

\setlength{\oddsidemargin}{0.3in}
\setlength{\evensidemargin}{-0.3in}

{\center
%\begin{center}
\vglue .08in

{\large\bf  \title }  \\[.5in]

\author \\[0.3in]

${}^{\small \mbox\Scorpio}${\em
Departamento de Ciencias F\'isicas,\\
Facultad de Ciencias Exactas,\\
Universidad Andres Bello,\\
Santiago de Chile.}
\\[0.1in]
and
\\[0.1in]
%\vskip 0.2in
${}^{\small \mbox\Pisces, \mbox\Virgo}${\em
Center for String and Particle Theory\\
Department of Physics,\\
University of Maryland at College Park,\\
College Park, MD 20742-4111 USA.}
\\[0.5cm]

{\bf ABSTRACT}\\[.01in]
%\end{center}
}

\begin{quotation}
%\abstract{
We investigate the complex of differential forms in curved, six-dimensional, $N=(1,0)$ superspace. The superconformal group acts on this complex by super-Weyl transformations. An ambi-twistor-like representation of a second conformal group arises on a pure spinor subspace of the cotangent space. The $p$-forms are defined by super-Weyl-covariant tensor fields on this pure spinor subspace. The on-shell dynamics of such fields is superconformal. We construct the superspace de Rham complex by successively obstructing the closure of forms. We then extend the analysis to composite forms obtained by wedging together forms of lower degree 
Finally, we comment on applications to integration in curved superspace and give a superspace formulation of the abelian limit of the non-abelian tensor hierarchy of $N=(1,0)$ superconformal models and propose a formulation of it as a Chern-Simons theory on the ambi-twistor-like superspace. 
%}%End abstract

%email addresses
%\vspace*{-.5cm}
\begin{flushleft}
~\\
{${}^{\small \mbox\Scorpio}$\href{mailto:ce.arias@uandresbello.edu}{ce.arias@uandresbello.edu}}\\
%\makebox[0pt][t]
{${}^{\small \mbox\Pisces}$ \href{mailto:wdlinch3@gmail.com}{wdlinch3@gmail.com}}\\
%\makebox[0pt][t]
{${}^{\small \mbox\Virgo}$ \href{mailto:alecridgway@gmail.com}{alecridgway@gmail.com}}
\end{flushleft}

\endtitle

%%%%%%%%%%%%%%%%%%%%%%%%%%%%%%%%%%%%%%%%%%%%%%%%%%%%%%%%%%%%%%%%%%
%%%%%%%%%%%%%%%%%%%%%%%%%%%%%%%%%%%%%%%%%%%%%%%%%%%%%%%%%%%%%%%%%%
\section{Introduction}
The investigation of the structure of superforms is an important step toward understanding the geometry of the underlying superspace. Due to the interplay between the spinors and tensors in such spaces, this structure is non-trivial even in the flat case. In the flat, four-dimensional, $N=1$ superspace, this is textbook material (see \S 4.4 of {\em Superspace} \cite{Gates:1983nr} and the original ref. \cite{Gates:1980ay}). 

In flat, four-dimensional, $N=2$ harmonic superspace a systematic analysis was carried out by Biswas and Siegel \cite{Biswas:2001wu}. 
%In curved superspaces or in flat superspace of  dimension higher than four, however, no systematic analysis has, to our knowledge, been reported.
In curved superspaces or in flat superspace of dimensions other than four, 
partial results may be found throughout the literature, with interesting developments in four and five dimensions being reported as recently as last year \cite{Kuzenko:2013rna, {Bossard:2013rza}, Butter:2012ze}. %\footnote
{Systematic studies of the closely related integral invariants in various dimensions are being carried out by Howe and his collaborators ({\em e.g.} ref. \cite{Bossard:2012xs, {Greitz:2011da}, Bossard:2010pk}).} Many of these results have been extended to three dimensions even more recently by Kuzenko and collaborators (see {\em e.g.} ref. \cite{Kuzenko}).
%\footnote
{In ten dimensions, composite forms and their couplings to supergravity in superspace were used early on to construct effective actions for massless string states ({\it e.g.} ref. \cite{Jim}). With the advent of covariant superstrings, a systematic analysis for the forms arising in the pure spinor superspace was performed by Berkovits and Howe \cite{Berkovits:2008qw} (see also \cite{Greitz:2011da}).} 

A special place in the hierarchy of superspaces is occupied by six-dimensional, $N=(1,0)$ Minkowski space as it has the largest isometry group compatible with the existence of eight real supercharges.
%the off-shell closure of the Poincar\'e algebra \cite{??}. 
The auxiliary field problem (that there are strictly more fermionic auxiliary fields required for off-shell closure than bosonic ones \cite{Siegel:1981dx}) is solved by retaining an $S^2\simeq SU(2)_R/U(1)_R$ part of the $R$-symmetry group in the quotient construction \cite{harm}. Closely related to this ``harmonic'' superspace is the ``projective'' superspace of \cite{projective}.
The extension of this to curved, six-dimensional superspace was presented in \cite{Linch:2012zh} where certain super-Weyl-covariant field representations were defined and an action principle was proposed.\footnote{This extension to six dimensions is based on previous work in 
five dimensions \cite{5D, {Kuzenko:2007hu}} and below \cite{234D}.}%end footnote

In this work, we continue the exploration of this superspace by analyzing the structure of the complex of differential forms. Motivated by its interpretation as the target space for covariant heterotic strings compactified on K3 \cite{oai:arXiv.org:1109.3200}, we introduce formal variables generating an algebra isomorphic to the graded exterior algebra in section \ref{S:Superforms}. This algebra has a subalgebra corresponding to the projection of a bosonic spinor $s\mapsto \lambda \otimes v$ to the product of a pure spinor $\lambda$ and iso-twistor variable $v$. We show that the differential forms are naturally defined on this subalgebra. In section \ref{S:Conformal}, we exploit the existence of the pure spinor subspace of differential forms to define super-Weyl covariant tensor fields and construct an ambi-twistor-like representation of the six-dimensional conformal group that acts on these tensor fields. The resulting {\em superconformal} field representations are subsequently used in section \ref{S:deRham} to derive the superspace analogue of the de Rham complex. 

In section \ref{S:Composite} we construct composite forms by wedging the various forms from section \ref{S:deRham}. 
In doing so, we are able to give a geometrical interpretation to certain multiplets constructed in \cite{Linch:2012zh}. 
Finally, we comment on some applications of the formalism in section \ref{S:Applications}. One of these is a reformulation of the non-abelian tensor hierarchy \cite{Samtleben:2011fj, Samtleben:2012mi}, reviewed in appendix \ref{S:NonAbTensorHierarchy}, in terms of a Chern-Simons theory on the ambi-twistor-like superspace of section \ref{S:Conformal}. Additionally, since the Bianchi identity for the differential forms were solved in a curved background, this automatically determines the couplings of the abelian part of the tensor hierarchy to supergravity.
Appendix \ref{S:6DGeom} summarizes some results of reference \cite{Linch:2012zh} on curved, six-dimensional, $N=(1,0)$ superspace. 

\paragraph{A note on notation} After careful consideration, we have decided to present our analysis (mostly) in an index-free notation because we are convinced that the benefits of doing so outweigh the risks. %Although this introduces the risk of alienating some readers, there are many more benefits.
Firstly, this presentation most closely resembles our method of calculation and discovery and is useful in proofs. Secondly, the notation simplifies comparison to the covariant string theories and higher gauge theories to which it is closely related. Thirdly, we will need a representation of the conformal group defined in terms of these variables in order to build a certain differential complex of superfields in terms of which the $p$-form components are defined. Finally, the presentation of the resulting complex is more easily compared to the results on higher gauge theory existing in the literature. Conversely, using the more familiar, ordinary superspace notation would significantly complicate substantial parts of the presentation. A standard analysis of each superform will be presented in \cite{CesarThesis}.

%%%%%%%%%%%%%%%%%%%%%%%%%%%%%%%%%%%%%%%%%%%%%%%%%%%%%%%%%%%%%%%%%%
%%%%%%%%%%%%%%%%%%%%%%%%%%%%%%%%%%%%%%%%%%%%%%%%%%%%%%%%%%%%%%%%%%
\section{Closed Superforms}
\label{S:Superforms}
In this section we write down the Bianchi identities for $p$-forms in six dimensions. We introduce some notation to simplify the calculations and find that the forms are supported on a certain subspace of the full cotangent space. Some results from reference \cite{Linch:2012zh} used in our calculations are summarized in appendix \ref{S:6DGeom}.

A $p$-form $\omega$ is closed iff
\begin{align}
d \omega = 0
\end{align}
where $d$ is given locally by $d\theta^{\alpha i} \partial_{\alpha i} + dx^a \partial_a$. For notational convenience, we will replace the basis forms $d\theta^{\alpha i}$ and $dx^a$ by new variables $s^{\alpha i}$ and $\psi^a$. To automatically incorporate the super-anti-commutative nature of the wedge product, we take the $\{s^{\alpha i}\}$ to be bosonic spinor variables and the $\{\psi^a\}$ to be fermionic vector variables
\begin{align}
s^{\alpha i} s^{\beta j} = s^{\beta j} s^{\alpha i} 
	,~~
	s^{\alpha i} \psi^a = \psi^a s^{\beta j}
	,~~
	\psi^a \psi^b=-\psi^b \psi^a .
\end{align}

The super-$p$-form $\omega=\sum_{s=0}^p \omega_{(s, p-s)}$ splits up as a sum of super-$s|(p-s)$-forms\footnote{
We are borrowing a compact, index-free notation from the theory of ordinary tensor fields on manifolds in which the subscript indicates the vector field on which the corresponding index is contracted ({\it e.g.} $\nabla_X$ for the directional derivative along the vector field $X$).}
\begin{align}
\omega_{(s, p-s)} := 
\omega_{s\dots s\psi\dots \psi} := 
	s^{\alpha_1 i_1}\dots s^{\alpha_s i_s} \psi^{a_1}\dots \psi^{a_{p-s}} \omega_{\alpha_1 i_1 \dots \alpha_s i_s a_1\dots a_{p-s} }, 
\end{align}
where the $p$-form components $ \omega_{\alpha_1 i_1 \dots \alpha_s i_s a_1\dots a_{p-s} }=  \omega_{\alpha_1 i_1 \dots \alpha_s i_s a_1\dots a_{p-s} }(x,\theta)$ are ordinary $N=(1,0)$ superfields.
The differential $d$ splits into two differentials $\partial_s$ and $\partial_\psi$
\begin{align}
d=\partial_s + \partial_\psi 
	:= s^{\alpha i} \partial_{\alpha i} + \psi^a \partial_a
	:= s^{\alpha i} {\partial \over \partial \theta^{\alpha i}} + \psi^a {\partial \over \partial x^a},
\end{align}
and the closure condition fans out into a collection of conditions respecting the grading by number of $s^{\alpha i}$-type variables. Thus, the closure condition may be represented compactly as
\begin{align}
s!(d\omega)_{ss\dots \psi\psi}=s \partial_s \omega_{s\dots \psi\psi\dots} + (-1)^s (p+1-s) \partial_\psi \omega_{ss\dots \psi\dots}  = 0
\end{align}
with the $s$ in front of the first term denoting the total number of $s$s in that term and with $p$ denoting the total degree of the $p$-form $\omega$.\footnote{Whenever we use the letter $s$ for a number, we mean the number of $s^{\alpha i}$-type variables appearing in the relevant formula. For example, the component $s!(d\omega)_{\alpha i \beta j \gamma k de}$ for a 4-form $\omega$ has $s=3$ and $p=4$ giving $3 \partial_s \omega_{ss\psi \psi} -2 \partial_\psi \omega_{sss\psi}$. We will simplify such formul\ae{} by multiplying by symmetry factors to cancel denominators, as we have done here.}

The differentials $\partial_s$ and $\partial_\psi$ do not commute with supersymmetry transformations. 
To get supersymmetrically covariant $p$-form components, we must replace coordinate derivatives by supercovariant derivatives.
In terms of flat supercovariant derivatives $D_s$ and $D_\psi=\partial_\psi$, the closure condition acquires a flat-space torsion term $T^a_{ss} = 2 i {s^i \gamma^a s_i}= s^{\alpha i } (\gamma^a)_{\alpha \beta} s^\beta_i $ :
\begin{align}
s D_s \omega_{s\dots s\psi \dots \psi} + (-1)^s (p+1-s) \partial_\psi \omega_{s\dots s \psi\dots \psi}  
	-  i (-1)^s s(s-1) \omega_{s\dots s\gamma(s,s)\psi\dots \psi} = 0,
\end{align}
where the notation $\omega_{\gamma(s,s) \dots}$ is shorthand for the contraction of the null vector $\gamma^a(s, s):=s^i\gamma^as_i$ on the component $\omega_{a \dots}$ (c.f. \ref{E:gammaomega}). 
In the curved superspace version of this closure relation, there are additional torsion terms
\begin{align}
\label{E:BIGeneral}
s \mathcal D_s \omega_{s\dots s\psi \dots \psi} + (-1)^s (p+1-s) \mathcal D_\psi \omega_{s\dots s \psi\dots \psi}  -
	 i (-1)^s s(s-1) \omega_{s\dots s\gamma(s,s)\psi\dots \psi}\cr
	+(-1)^s s (p+1-s) T_{s\psi}{}^{\alpha i} \omega_{\alpha i s\dots s \psi\dots \psi}
	- \tfrac12 (p+1-s)(p-s) T_{\psi \psi}{}^{\alpha i} \omega_{\alpha i s\dots s \psi\dots \psi}
=0.
\end{align}
These torsions, and the six-dimensional curved supergeometry in general, are reviewed in appendix \ref{S:6DGeom}. In section \ref{S:deRham} we will be solving this equation for each $p=1,\dots, 6$.

Superforms with no restrictions on their components yield {\em reducible} representations of the super-Poincar\'e algebra.
Irreducible representations result by setting to zero one irreducible component. Once this is done, the closure conditions become non-trivial consistency conditions on the other components ({\it e.g.} all components of lower dimension are required to vanish). Following common superspace practice, we will refer to the resulting closure conditions as ``Bianchi identities''. 
%procedure of solving the set of closure conditions as solving the ``Bianchi identities''.
Thus, to begin solving the Bianchi identities of any particular $p$-form, we must locate the component of lowest engineering dimension that does not vanish.

Let $n$ denote the largest value of $s$ such that the $s|(p-s)$-component of the $p$-form $\omega$ is non-vanishing. Then equation (\ref{E:BIGeneral}) simplifies to
\begin{align}
\label{E:BILowest}
(n+1) \mathcal D_s \omega_{\scriptsize{\underbrace{s\dots s}_n \underbrace{\psi \dots \psi}_{p-n}}} =  i (-1)^{s+1} n(n+1) \omega_{s\dots s\gamma(s,s)\psi\dots \psi}.
\end{align}
If we further take the projection $s^{\alpha i}\mapsto \lambda^\alpha v^i$ to the product of an unconstrained bosonic spinor $\lambda$ of $SU(4)$ and an unconstrained bosonic spinor $v$ of $SU(2)$, then the right-hand side vanishes and we find that 
\begin{align}
\label{E:Qomega}
Q \omega_{\scriptsize{\underbrace{\lambda\dots \lambda}_n \underbrace{\psi \dots \psi}_{p-n}}}  = 0.
\end{align}
Here,
\begin{align}
Q :=\mathcal  D_{\lambda \otimes v} = \lambda^\alpha v^i \mathcal D_{\alpha i} =:  \lambda^\alpha \mathcal D^+_{\alpha}
\end{align}
stands for the projected superspace derivative. 

The operator $Q$ 
%This operator 
appears repeatedly in the analysis that follows. Its importance derives from the fact that the condition (\ref{E:Qomega}) holds for the lowest non-vanishing component of any $p$-form and, as such, it universally appears as a defining condition on the superfield from which all other components are derived. 
We call the projection $s \mapsto \lambda\otimes v$ the {\em pure spinor projection} for reasons discussed in appendix \ref{S:6DGeom} (c.f. eq. (\ref{E:PureSpinor}) and the surrounding discussion). 

Superfields $\Phi = \Phi(x, \theta, v)$ satisfying $\mathcal D^+_{\alpha} \Phi =0$ are known in the projective superspace literature as {\em analytic} superfields.\footnote{We are glossing over some subtleties here which we address as needed in the sequel.} When $\Phi$ is homogeneous of degree $n$ in the variables $v$, $\Phi$ is said to have {\em homogeneity weight} $n$. When we wish to indicate this explicitly, we will do so with a superscript $\Phi^{+n}$. In these terms, equation (\ref{E:Qomega}) says that the pure spinor projection $\omega^{+n}_{\lambda\dots \lambda \psi\dots \psi}$ of the lowest-dimension, non-vanishing component of the $p$-form is an analytic superfield with homogeneity weight $n$. The dimension of this field is $d= \tfrac12 n + p- n = p-\tfrac n2$. 

It is possible for the aforementioned projection of the $n|(p-n)$-component to vanish.\footnote{This happens, for example, for the $p=3$ form $H$ where $n=2$ ({\it i.e.} $H_{ss\psi}\neq 0$) but $H_{\lambda \lambda \psi} \equiv 0$ (c.f. \S\ref{S:3formFS}).} When this happens, the pure spinor projection of the lowest  dimension ($d+\tfrac12$) Bianchi identity is trivially satisfied. Passing to the next Bianchi identity (that with dimension $d+1$) we find that it is the $(n-1)|(p-n+1)$-component of $\omega$ that projects to an analytic superfield provided this projection does not also vanish identically. If it does, we pass to the next Bianchi identity. We proceed this way until we find a value $n^\prime\leq n$ such that the projection of the $n^\prime|(p-n^\prime)$-component does not vanish under pure spinor projection and, therefore, defines an analytic superfield with homogeneity weight $n^\prime$.\footnote{In the $p=3$ example, $n=2$ but $n^\prime =1$ (c.f. \S\ref{S:3formFS}).} 
In our analysis, we will find that this field is a {\em superconformal primary field}, that is, a superfield transforming homogeneously under super-Weyl transformations as we recall in section \ref{S:Weyl} (c.f. eq. \ref{E:WeylWeight}). 

In section \ref{S:Weyl}, we will also show that $Q^2=0$ on superfields defined over the pure spinor subspace. These superfields, graded by homogeneity weight and spin, form a complex of spaces with differential $Q$. Assuming this, we conclude that differential forms restricted to the pure spinor subspace are sourced by superconformal primary superfields in the cohomology of this complex. In the next section,
%(\S\ref{S:Conformal}) 
we will study large families of such superfields. 

With the lowest non-vanishing component in hand, the remaining components of the superform can be reconstructed by the usual method. To wit, one first solves the lowest non-trivial Bianchi identity for the dimension-$(d+\tfrac12)$ component by inverting the constant torsion $2i (s^i\gamma^as_i)$ in equation (\ref{E:BILowest}). These components suffice to solve the next Bianchi identity (\ref{E:BIGeneral}) for the dimension-$(d+1)$ component, provided certain additional constraints are imposed on the defining field. This process continues to define the next-higher component and, in principle, additional constraints, until we have reached the top component of dimension $p$. The final two Bianchi identities must now be identically satisfied. In section \ref{S:deRham}, we will carry out this procedure to find the components of all the differential forms in curved superspace and verify that the final two identities are satisfied identically in the flat limit.

%%%%%%%%%%%%%%%%%%%%%%%%%%%%%%%%%%%%%%%%%%%%%%%%%%%%%%%%%%%%%%%%%%
%%%%%%%%%%%%%%%%%%%%%%%%%%%%%%%%%%%%%%%%%%%%%%%%%%%%%%%%%%%%%%%%%%
\section{Conformal symmetry}
\label{S:Conformal}
Consider a superfield $\Phi^{\beta_1\dots \beta_c}{}_{\alpha_1\dots \alpha_s k_1\dots f_f}(x, \theta)$ with $c$ symmetric fundamental spinor indices, $s$ symmetric anti-fundamental spinor indices, and $f$ symmetric isospin indices. We introduce the commuting variables $\{\lambda^\alpha, \bar \lambda_\alpha, v^i\}$ in the $\{(\mathbf 4, \mathbf 1), (\bar{\mathbf 4}, \mathbf 1), (\mathbf 1, \mathbf 2)\}$ representations of $SU(4)\times SU(2)$ and replace the superfield with 
\begin{align}
\label{E:Family}
\Phi_{\scriptsize{\underbrace{\lambda \dots \lambda}_s}  \scriptsize{\underbrace{\bar \lambda \dots \bar \lambda}_c} }^{+f}=
	v^{i_1}\dots v^{i_f} 
	\lambda^{\alpha_1}\dots \lambda^{\alpha_s} 
	\bar \lambda_{\beta_1}\dots \bar \lambda_{\beta_c}
	\Phi^{\beta_1\dots \beta_c}{}_{\alpha_1\dots \alpha_s i_1\dots i_f} .
\end{align}
Lorentz-irreducibility requires tracelessness on pairs of fundamental and anti-fundamental spinor indices. We impose this by requiring
\begin{align}
\label{PSconstraint}
\lambda^\alpha \bar \lambda_\alpha = 0.
\end{align}

Introducing the conjugate momenta $\{\omega_\alpha, \bar \omega^\alpha, p_i\}$ allows us to rewrite the action of the Lorentz and isospin generators as
\begin{align}
\label{E:MJRep}
 \mathfrak m_{ab} =  -\tfrac12 \lambda \gamma_{ab} \omega  - \tfrac 12 \bar \lambda \tilde \gamma_{ab}\bar \omega 
~~~\mathrm{and}~~~
 \mathbf j_{ij} =  v_{(i} p_{j)} .
\end{align}
This representation preserves the constraint (\ref{PSconstraint}). The other derivations preserving the constraint are 
\begin{align}
\label{E:ConfGen2}
 \mathfrak p_a = \bar \lambda \tilde \gamma^a \omega ,~~
 \mathfrak k^a = \lambda \gamma^a \bar \omega ,~~
\Delta = \tfrac 12 \lambda^\alpha \omega_\alpha - \tfrac 12 \bar \lambda_\alpha \bar \omega^\alpha 
,~~\mathrm{and}~~
 \mathbf q= v^i p_i.
\end{align}
Together, they generate the conformal algebra $\mathfrak{so}_{6,2}$
\begin{align}
\label{E:so8*}
&[ \mathfrak m_{ab},  \mathfrak m_{cd}] = - 2\eta_{c[a}  \mathfrak m_{b]d} + 2\eta_{d[a}  \mathfrak m_{b]a} 
	,~[ \mathfrak p_a,  \mathfrak k^b] = 2 \delta_a^b \Delta -2  \mathfrak m_a{}^b  ,\cr
&[ \mathfrak m_{ab},  \mathfrak p_c] = - 2\eta_{c[a}  \mathfrak p_{b]} 
	,~ [\Delta,  \mathfrak p_a] = -  \mathfrak p_a ,\cr
&[ \mathfrak m_{ab},  \mathfrak k_c] =  -2\eta_{c[a}  \mathfrak k_{b]}
	,~ [\Delta,  \mathfrak k_a] =   \mathfrak k_a,
\end{align}
and a decoupled $\mathfrak u_1$.
%\footnote{\color{red} Check reference \cite{Govil:2014uwa}.}

An irreducible (iso-)spin-tensor (\ref{E:Family}) is an irreducible representation of this algebra. The derivation $Q$ acts on such representations. In general, its square is proportional to Lorentz and isospin generators (c.f. eq. \ref{E:Q^2}). When acting on the representations (\ref{E:Family}), however, the Lorentz $M\to \mathfrak m$ and isospin generators $J\to \mathbf j$ are represented by (\ref{E:MJRep}). Simple Fierz rearrangement then implies that the only remaining term is proportional to the constraint (\ref{PSconstraint}). Therefore, 
\begin{align}
Q^2=0 
\end{align}
when acting on the family of fields of the form (\ref{E:Family}). Therefore, these fields form a differential complex graded by spin and isospin.

%%%%%%%%%%%%%%%%%%%%%%%%%%%%%%%%%%%
\subsection{Superconformal primary superfields}
\label{S:Weyl}
In this subsection, we use the pure spinor/ambi-twistor-like representation just introduced to construct large families of superconformally covariant field representations by imposing super-Weyl-invariant constraints. The ambi-twistor-like variables are not necessary to define the constraints but they greatly simplify the proof of their super-Weyl covariance. When the resulting representations are on shell, their dynamics are superconformally invariant.

In appendix \ref{S:6DGeom} we recall the action of super-Weyl transformations (c.f. eq. \ref{E:WeylTransformation}--\ref{E:NWeylTransformation}) preserving the algebra of covariant derivatives defining the six-dimensional, $N=(1,0)$ supergravity theory studied in reference \cite{Linch:2012zh}. These transformations are parameterized by a real, unconstrained scalar superfield $\sigma(z)$. In keeping with superspace terminology, we define an irreducible superfield (\ref{E:Family}) to be a {\em Weyl tensor of weight $w$} provided that under such a transformation
\begin{align}
\delta \Phi = w \sigma \Phi.
\end{align}
On such fields, the differential transforms as
\begin{align}
\delta Q &= \tfrac12 \sigma Q  +  \tfrac12 \left(3s+c-4f\right) (Q\sigma ) ,
\end{align}
as follows from (\ref{E:WeylTransformation}) evaluated on the representation (\ref{E:MJRep}).
Therefore, if $\Phi$ is a Weyl tensor with weight $w$ then $Q\Phi$ will be a Weyl tensor of weight $w+\tfrac12$ provided\footnote{The parameter $\mathcal D_{\alpha i} \sigma$ is the $S$-supersymmetry parameter. Canceling this term implies that the field $\Phi$ is invariant under $S$-supersymmetry, that is, it is a superconformal primary field.}
\begin{align}
\label{E:WeylWeight}
w = 2f - \tfrac32s - \tfrac12 c .
\end{align}
In particular, it is consistent to constrain 
\begin{align}
\label{E:Q=0}
Q\Phi =0
~~~\mathrm{provided}~~~
w = 2f - \tfrac32s - \tfrac12 c .
\end{align}

Important examples are given in the following table:
\begin{align}
\begin{array}{c|cccccc}
p	& \Phi & (f,s,c) & \mathrm{superfield} & \mathrm{field~strength/potential} & w & \mathrm{section} \\ \hline
1 	& 	A	&	(1,1,0) &	A_{\alpha i} 	&	\mathrm{potential}	&	1/2	&\ref{S:1formFS}\\
2 	& 	W	&	(1,0,1) &	W^{\alpha i}	&	\mathrm{field~strength} 		& 	3/2	&\ref{S:2formFS}\\
2 	&	V	&  (1,0,1) &	V^{\alpha i}	&	\mathrm{potential} 	& 	3/2 	&\ref{S:3FormAlternative}\\
3 	&	C	&  (2,1,1) &	C_{ab ij}	&	\mathrm{potential} 	& 	2 	&\ref{S:4FormAlternative}\\
5	&  K	&	(2,0,0)	&	K_{ij}				&	\mathrm{field~strength}		&	4	&\ref{S:5formFS}\\
6	&  L	&	(3,1,0)	&	L_{\alpha ijk}	&	\mathrm{field~strength}		&	9/2	&\ref{S:6formFS}
\end{array}
\nonumber
\end{align}
As we will discuss in detail in section \ref{S:deRham}, they represent, respectively, the gauge 1-form potential $A:=\lambda^\alpha v^i A_{\alpha i}$, its 2-form field strength $W:= \bar \lambda_\alpha v_i W^{\alpha i}$, the gauge 2-form potential $V:=\bar \lambda_\alpha v_i V^{\alpha i}$, the gauge 3-form potential ${C}:= (\lambda \gamma^{ab} \bar \lambda)  v^iv^j \, {C}_{ab\, ij}$, the so-called {\em linear multiplet} $K := v^i v^j K_{ij}$ related to the projective Lagrangian density,\footnote{The projective Lagrangian density \cite{Linch:2012zh} is a homogeneity weight-2, analytic superfield $\hat K^{++}(\zeta)$ containing an infinite number of ordinary superfields. Of these, $K_{ij}$ are the first three terms in an expansion in $\zeta:=v^2/v^1$ with all other superfields containing only auxiliary components.} and the 6-form field strength $L:= \lambda^\alpha v^i v^ j v^k L_{\alpha ijk}$.
%, as we will discuss in detail in section \ref{S:deRham}.

In addition to this family of representations, there is an infinite family of symmetric, traceless ``spin-$\ell$'' superfields $J_{c_1\dots c_\ell i_1\dots i_f}=J_{(c_1\dots c_\ell) (i_1\dots i_f)} -\mathrm{traces}$: Let
\begin{align}
J^{(\ell)}_f := v^{i_1} \dots v^{i_f} k^{c_1} \dots k^{c_\ell} J_{c_1\dots c_\ell i_1\dots i_f} .
\end{align}
Then the condition 
\begin{align}
\label{E:HigherSpin}
QJ^{(\ell)}_f = 0 ~~~\Rightarrow~~~ w = 2f-\ell ,
\end{align}
defines a Weyl superfield of weight $2f-\ell$. We will see section \ref{S:4formFS} that for $\ell =1$ and $f=2$, this condition defines the weight $w=3$ field $G_{a\, ij}$ sourcing the 4-form field strength. 

Finally, there are seven other families of Weyl superfields that are described naturally in various alternative ``polarizations'' of the pure spinor variables. To illustrate what we mean by this, we will explicitly work out the only example used in our analysis of differential forms: Consider a superfield with $s=0$ but $c$ and $f$ arbitrary. Performing the canonical transformation from $\{\bar\lambda, v\}$ variables to $\{\omega, p\}$ on the superfield $\Phi \mapsto \Phi^\dagger$ only, results in a re-``normal''-ordering under which the conformal weight changes as $w = 2 v^ip_i -\tfrac12 \bar \lambda_\alpha \bar \omega^\alpha -\tfrac 32 \lambda^\alpha \omega_\alpha \mapsto 2[-f-2]-\tfrac 12 [0]-\tfrac32[-c-4]$. In the new polarization, the constraint $Q\Phi^\dagger = 0$ is equivalent to the condition\footnote{Equivalently, we can keep $\Phi$ and perform the transformation on $Q\mapsto Q^\dagger$. Then $Q^\dagger$ will act by contraction of the form $\Phi$ on the vector $\mathcal D_{\lambda \otimes v}$. }
\begin{align}
\label{E:DivConstraint}
\mathcal D_{\alpha i} \Phi^{\alpha \beta_1\dots \beta_{c-1}\, i j_1\dots j_{f-1}} =0 
	~~~\Rightarrow~~~ w = -2f +\tfrac 32 c+2
\end{align}
on the {\em contraction} of the operator $\mathcal D_{\alpha i}$ with the spin and isospin indices on $\Phi^{\alpha i\dots}$.
This constraint is compatible with the condition (\ref{E:Q=0}) when the weights agree, that is, for $c=2f-1$ and $w=f+\tfrac12$. As we will derive in section \ref{S:2formFS}, the lowest-weight member of this tower is the the gauge 1-form field strength superfield $W$. Note that we are not {\em required} to impose this condition on such a weight-$(f+\tfrac12)$ field. If we do not, we find that the component $Q\Phi^\dagger$ is another Weyl tensor of weight $f+1$. We will see in section \ref{S:3FormAlternative} that this observation provides the link between the description of the 2-form gauge potential described by the supefield $V$ and that in terms of its 3-form field strength. The latter is built on the Weyl-weight-2, real scalar superfield $\Phi$ related to $V$ by $\Phi = \mathcal D_{\alpha i} V^{\alpha i}$ (c.f. eq. \ref{E:PhiDefinition}).

%%%%%%%%%%%%%%%%%%%%%%%%%%%%%%%%%%%%%%%%%%%%%%%%%%%%%%%%%%%%%%%%%%
%%%%%%%%%%%%%%%%%%%%%%%%%%%%%%%%%%%%%%%%%%%%%%%%%%%%%%%%%%%%%%%%%%
\section{The super-differential complex}
\label{S:deRham}
In this section, we will explicitly go through the steps outlined at the end of section \ref{S:Superforms} for solving the closure conditions (referred to as Bianchi identities) for $p$-forms with $p=1,\dots, 6$ subject to the condition that certain components vanish ({\it e.g.} $F_{ss}=0$, $H_{sss}=0$). In the process, we will find that for each $p\leq 5$, there is an additional constraint on the defining superfield necessary for the closure of the dimension-$p$ Bianchi identity. To pass to the next $p$-form in the complex, we relax this last condition, thereby obstructing the closure of the $p$-form field strength. In doing so, we find the defining superfield for the $(p+1)$-form. In this sense, {\em the entire complex is derived from the constraint $F_{\alpha i\beta j}=0$ imposed on the 2-form}. The lowest non-vanishing components of the resulting forms have precisely the dimensions found to imply superconformal invariance in section \ref{S:Weyl}.

The result for flat forms with $p=2,\dots, 5$ is represented schematically in the following table: 

{\scriptsize
\begin{align}
\begin{array}{llll}
~~~p=2 & ~~~~p=3& ~~~~~p=4& ~~~~~~p=5\\
\\ 
F_{ss} = 0 \\
F_{s\psi} = is^i\gamma_\psi W_i & H_{sss} =0 \\
F_{\psi\psi} = D^i\tilde \gamma_{\psi\psi}W_i & H_{ss\psi} =s^i\gamma_\psi s_i \Phi & G_{ssss} = 0 \\
	&H_{s\psi\psi} = s^i\tilde \gamma_{\psi\psi}D_i \Phi& G_{sss\psi} =0  & K_{sssss} =0 \\
	& H_{\psi\psi\psi} =D^i\tilde \gamma_{\psi\psi\psi}D_i \Phi& G_{ss\psi\psi} = s^i\gamma_{\psi\psi a}s^jG^a_{ij} & K_{ssss\psi} =0 \\
		&&G_{s\psi\psi\psi}= *s^i\gamma_{\psi\psi}D^jG_{\psi ij} & K_{sss\psi\psi}=0 \\
		&& G_{\psi\psi\psi\psi}= * D_i\tilde \gamma_\psi D_j G_{\psi}^{ ij}  & K_{ss\psi\psi\psi} = s^i\gamma_{\psi\psi \psi}s^jK_{ij}  \\
			&&& K_{s\psi\psi\psi\psi}= *s^i\gamma_{\psi\psi} D^j K_{ij} \\
			&&& K_{\psi\psi\psi\psi\psi} =* D_i\tilde \gamma_\psi D_j K^{ij} 
\\ \\ \hline \\
D_{(i}\tilde \gamma_{ab} W_{j)} \stackrel{(\ref{E:XCondition1})}= 0 
	&&\Pi_a^b D_{(k} G_{b ij)} \stackrel{(\ref{E:GConstraint1})}=0 
	& D_{(k} K_{ij)} \stackrel{(\ref{E:5formConstraint})}= 0\\
D_iW^i \stackrel{(\ref{E:PhiCondition})}=0 
	& 	D^2_{a\,ij} \Phi \stackrel{(\ref{E:DDPhiConstraint})}= 0 
	& D^{2a}_{k(i} G_{ai)}{}^k \stackrel{(\ref{E:GConstraint2})}=i\partial^a G_{a ij} 
\end{array}
\nonumber
\end{align}}
In order to fit the entries into the table, we have suppressed the 1- and 6-form, are ignoring numerical factors, and we use $*$ in the higher components of the forms of degree $p=4,5$ to schematically denote the Hodge dual. 

Very roughly, going up a $p$-form chain corresponds to applying the operator $D_i \tilde \gamma_\psi \partial/\partial s_i$. Going across corresponds to finding the dimension-$p$ component and replacing the $DD$ operator acting on the defining field with some field that is bilinear in $s$. In fact, this is happening because the Bianchi identity has {\em reducible} Lorentz/isospin components while the $DD$ field strength of the $p$ form is irreducible. Imposing that the additional irreducible components vanish closes the $p$-form Bianchi identity $d\omega_{p}=0$. Alternatively, we can interpret this additional irreducible part as the lowest non-trivial component of a $(p+1)$-form $\omega_{p+1}$. Then the statement is that the non-vanishing of this new form is the obstruction to the $p$-form Bianchi identity, that is, $d\omega_{p}=\omega_{p+1}$. 

For example, starting with $F_{ss}=0$ and working our way up to the Bianchi identity $(dF)_{ss\psi} = 0$, we find that $F_{\psi\psi} \propto D^i\tilde \gamma_{\psi\psi}W_i$ for the top component of the 2-form. However, in that same identity, there remains uncanceled the Lorentz-irreducible term $D_{\alpha i} W^{\alpha i}$ which the Bianchi identity sets to zero. Alternatively, we may decide to deform the closure condition by introducing a source $H_{ss\psi}$ at this level. Then the new identity is $dF = H$. Consistency requires $dH=0$, which we then proceed to solve. But this is just the Bianchi identity for the 3-form as it appears in the second column of the table.

Below the dividing line in the table are the conditions the defining superfields satisfy. The top line represents the relation $Q\Phi=0$ on the pure spinor cone. The line below it denotes additional conditions required for the closure of the Bianchi identity for that particular $p$-form. The interpretation of the $(p+1)$-form as an obstruction to the $p$-form Bianchi identity is reflected in the fact that the left-hand side of each condition on the bottom line is in the same irreducible Lorentz and isospin representation as the defining field to the right of it.\footnote{For example, for $p=2$ the spinor field strength has no scalar component $D_i W^i$: This term appears in the Bianchi identity at the same level as $F_{\psi\psi}$ but cannot be canceled by it since there is no Lorentz-invariant way to absorb a scalar into a 2-form. This component must therefore be set to vanish if we want the Bianchi identity for the 2-form to be satisfied. This scalar superfield is of the same form as the defining field $\Phi$ in the 3-form to the right of it in the table.} %end footnote
In the following subsections, we will make all of these statements explicit. 
%In the process, we will confirm that the superfields defining the various $p$-forms satisfy superconformally-invariant constraints.

We should mention that we are not claiming that the set of $p$-form representations we obtain is complete;\footnote{We thank Gabriele Tartaglino-Mazzucchelli for raising this question.} in lower dimensions, it is possible to have so-called ``variant representations'' \cite{{Gates:1980az}, Gates:1983nr, {Biswas:2001wu}}. However, the tower we obtain is uniquely determined by working our way up from the constraint $F_{ss}=0$. Furthermore, each form has a superconformal primary field as its lowest non-vanishing component. Taken together, it may be that the resulting complex is unique. Proving this or finding counter-examples (variant representations) should be possible by first classifying all superconformal constraints of scaling weight $w\leq p$ along the lines of section \ref{S:Weyl} and then inspecting them for proper Lorentz and iso-spin structure.\footnote{For example, $H_{sss}$ has scaling weight $\tfrac32$. There is a representation of this dimension which we call $V^{\alpha i}$ but it cannot be used to construct this particular component $H_{sss}$ since, for example, $s^{\alpha i} s^{\beta}_i s^{\gamma j} \varepsilon_{\alpha \beta \gamma \delta} V^{\delta}_j \equiv 0$. Instead, it enters into $H_{ss\psi}$ as $\mathcal D_{\alpha i} V^{\alpha i}$, as explained in section \ref{S:3FormAlternative}.}

%%%%%%%%%%%%%%%%%%%%%%%%%%%%%%%%
\subsection{The closed $1$-form}
\label{S:1formFS}
The components of a generic 1-form are $(A_s, A_\psi)$. The first closure condition for a closed 1-form is 
\begin{align}
\label{E:dAss}
2(dA)_{ss} = 2\mathcal D_s A_s - 2 i A_{\gamma(s,s)} = 0.
\end{align}
The pure spinor projection of this equation is simply
\begin{align}
\label{E:QA}
QA=0
\end{align}
where $A=\lambda^\alpha v^i A_{\alpha i}$ defines a Weyl tensor superfield provided $w=\tfrac12$. This condition, which is equivalent to $\mathcal D_{(\alpha (i} A_{\beta)j)} = 0$, was solved in flat space in references \cite{Howe:1983fr} and \cite{Koller:1982cs} based on the four-dimensional, $N=2$ solution of Mezincescu \cite{Mezincescu:1979af} as $A_{\alpha i }  = D_{\alpha i} {U}  + D_{\alpha}^j U_{ij}$. Substituting this back into the pure spinor constraint $QA=0$, we find that the isotriplet prepotential $U_{ij}$ is required to satisfy $D^k \tilde \gamma_{abc} D_k U_{ij} = 0$. Due to the flat-space identity $D^m \tilde \gamma_{abc} D_m D^4_{ijkl} \equiv 0$, this implies that it can be written in terms of the unconstrained Mezincescu prepotential $u^{ij}$ as $U_{ij} = D^4_{ijkl} u^{kl}$. 

Returning to curved space, the vector component the super-1-form is determined by the dimension-1 Bianchi identity to be 
\begin{align}
A_\psi = -\tfrac i 8 \mathcal D^k \tilde \gamma_\psi A_k.
\end{align}
Substituting the curved-space analogue of Koller's solution 
\begin{align}
\label{E:KollerSolution}
A_{\alpha i }  = \mathcal D_{\alpha i} {U}  +\mathcal D_{\alpha}^j U_{ij}
~~~\mathrm{with}~~~
\mathcal D^k \tilde \gamma_{abc} \mathcal D_k U_{ij}
	+256i N^{(-)}_{abc}U_{ij}=0, 
\end{align}
this becomes
$A_\psi = \mathcal D_\psi {U} -\tfrac i8 \mathcal D_{(i}\tilde \gamma_\psi \mathcal D_{j)} U^{ij}$.
The ${U}$ part drops out of the dimension-$\tfrac32$ Bianchi identity
\begin{align}
\label{E:ABI2}
\mathcal D_s A_\psi - \mathcal D_\psi A_s - T_{s\psi}{}^{\alpha i} A_{\alpha i}= 0
\end{align}
which constrains the prepotential $U_{ij}$ by $s_i \gamma_\psi W^i_U =0$ for
\begin{align}
W^{\alpha i}_U := \tfrac 18 \mathcal D^{3\alpha ijk} U_{jk} -\tfrac i6 (\tilde \gamma^a)^{\alpha \beta} \mathcal D_a \mathcal D_{\beta j} U^{ij} + \textrm{ torsion terms}.
\end{align}
Note that this is of the form of the field strength superfield for a {\em gauge} 1-form with prepotential $U_{ij}$. In the next section, we will relax the condition forcing it to vanish, thereby generating the closed 2-form field strength as an obstruction to the closure of the 1-form being worked out here.

Setting the field strength superfield $W_U=0$ is gauge equivalent to setting $U_{ij}=0$. Doing this, we find that the closed 1-form is given by $(A_s , A_\psi) = (\mathcal D_s {U}, \mathcal D_\psi {U})$. The dimension-2 Bianchi identity 
\begin{align}
2 \mathcal D_\psi A_\psi - \tfrac12 \cdot 2 T_{\psi \psi}{}^{\alpha i} A_{\alpha i} \equiv 0
\end{align}
is satisfied identically.
We conclude that, as expected, the unique closed, Weyl-covariant 1-form is the exact 1-form with weight-0 (iso-)scalar potential ${U} $.

%%%%%%%%%%%%%%%%%%%%%%%%%%%%%%%%
\subsection{The closed $2$-form}
\label{S:2formFS}
The closed 1-form $A$ of the previous section satisfied two constraints. The first of these was the pure spinor constraint $QA=0$ (\ref{E:QA}). The second was the dimension-$\tfrac32$ Bianchi identity (\ref{E:ABI2}) constraining the prepotential $U$ to vanish. We can deform this particular Bianchi identity by introducing the superfield 2-form field strength $F_{s\psi}=(dA)_{s\psi}$ as an obstruction to the closure of the 1-form field $A$. In this interpretation, continuing to impose the pure spinor condition (\ref{E:QA}) corresponds to keeping $F_{ss}=0$. 
%\footnote
{In fact, the two conditions are equivalent: Generally, $F_{\alpha i\beta j} =F_{\beta j\alpha i} = F_{(\alpha (i\beta) j)} + F_{[\alpha [i\beta] j]}$ but the second term is equivalent to $\varepsilon_{ij} (\gamma^a)_{\alpha \beta} F_a$ which can be absorbed by a field redefinition into the vector component $A_a$. The remaining term $F_{(\alpha (i\beta) j)} = \mathcal D_{(\alpha (i} A_{\beta) j)}$ is precisely the combination $QA$.} 

In this section, we will solve the Bianchi identities for the closed 2-form $F$ subject to 
\begin{align}
\label{E:Fss}
F_{ss}=0.
\end{align} 
We emphasize that this is the only input from section \ref{S:1formFS} that we will use.
The lowest-level Bianchi identity relates the components $F_{ss}$ and $F_{s\psi}$ as
\begin{align}
3\mathcal D_s F_{ss} +3\cdot 2 i F_{s\gamma(s,s)} =0.
\end{align}
By the Fierz identity (\ref{E:FundamentalFierz}), the condition $F_{ss}=0\Rightarrow  F_{s\gamma(s,s)}=0$ is compatible with the solution
\begin{align}
F_{s \psi} = 2i (s^i \gamma_\psi W_i) ,
	%= 2i \psi_{\gamma(s, W)}
\end{align}
for some positive chirality, fermionic superfield $W^{\alpha i}$. The dimension-2 identity reads
\begin{align}
\label{E:2FormBI2}
2\mathcal D_s F_{s\psi} +\mathcal D_\psi F_{ss} - 2 i F_{\gamma(s,s)\psi} =0.
\end{align}
The pure spinor projection of this implies 
\begin{align}
\label{E:XCondition1}
Q W = 0
\end{align}
in the sense of section \ref{S:Weyl}. To see this, note that under pure spinor projection, $(s^i \gamma_\psi W_i) \mapsto c_\alpha  W^{\alpha+}$ where we define $c_\alpha := (\lambda \gamma_\psi)_\alpha$. Although this contravariant spinor is fermionic, it satisfies $\lambda^\alpha c_\alpha \equiv 0$ analogously to (\ref{PSconstraint}) so that the analysis of that section remains valid in this special parameterization. The pure spinor condition (\ref{E:XCondition1}) defines a Weyl tensor superfield provided its weight is $w=\tfrac32$.
By condition (\ref{PSconstraint}) or, rather, its fermionic version, it is equivalent to 
\begin{align}
\label{E:XCondition2}
\mathcal D_{(i} \tilde \gamma_{ab} W_{j)} = 0.
\end{align}
As a practical matter, this condition says that at the level of component field strengths, there is no triplet of 2-forms.

The pure spinor projection isolates the isospin triplet part of the dimension-2 Bianchi identity (\ref{E:2FormBI2}). The remaining isospin singlet part contains a superfield 2-form term and a scalar term. Canceling the 2-form part results in the definition of the top component of the closed 2-form
\begin{align}
F_{\psi \psi} = -\tfrac 14 \mathcal D_k\tilde \gamma_{\psi\psi} W^k,
\end{align} 
leaving only $(s^i\gamma_\psi s_i) (\mathcal D_{\beta j} W^{\beta j}) = 0$ uncanceled. Therefore, in addition to the constraint (\ref{E:XCondition2}), we are required to impose the vanishing of the scalar term
\begin{align}
\label{E:PhiCondition}
\mathcal D_{\alpha i} W^{\alpha i} = 0.
\end{align}
Note that this condition is Weyl invariant by (\ref{E:DivConstraint}) since $w=\tfrac32$. Furthermore, it is required for consistency of that part of the dimension-2 Bianchi identity that drops out of the pure spinor projection. By contrast, the constraint (\ref{E:XCondition1}) is defined by the projection to the pure spinor subspace. In the next subsection (c.f. \S\ref{S:3formFS}), we will interpret the lowest 3-form component as an obstruction to this additional condition (\ref{E:PhiCondition}).

The constraints (\ref{E:XCondition2}) and (\ref{E:PhiCondition}) define the abelian 2-form field strength representation $W$ \cite{Siegel:1978yi}. Its derivative can be expanded as
\begin{align}
\mathcal D_{\alpha i} W^{\beta j} = \delta_i^j F_{\alpha}{}^{\beta} + \delta_\alpha^\beta X_i{}^j
\end{align}
for some iso-triplet superfield $X_{ij} = X_{(ij)}$ whose lowest components are auxiliary fields. 
The dimension-$\tfrac52$ Bianchi identity
\begin{align}
\mathcal D_s F_{\psi\psi} - 2 \mathcal D_\psi F_{s\psi} - 2 T_{s\psi}{}^{\alpha i} F_{\alpha i \psi} = 0
\end{align}
serves to relate the field equations to $X$ and imply no additional conditions on $W$. Likewise, the dimension-3 identity 
\begin{align}
\mathcal D_\psi F_{\psi\psi} - T_{\psi\psi}{}^{\alpha i} F_{\alpha i \psi} = 0
\end{align}
implies no new constraints. Consequently, this $(5+3)|(4+4)$-component representation is off-shell.

%%%%%%%%%%%%%%%%%%%%%%%%%%%%%%%%
\subsection{The closed $3$-form}
\label{S:3formFS}

The defining superfield $W$ of the abelian 2-form satisfies, in addition to the pure spinor condition (\ref{E:XCondition1}), the dimension-2 constraint (\ref{E:PhiCondition}) as follows from the $0=(dF)_{ss\psi} \propto (s^i\gamma_\psi s_i) (\mathcal D_{\beta j} W^{\beta j})$ Bianchi identity. In keeping with our general philosophy, we source this equation with
\begin{align}
	H_{ss\psi} = 2i(s^i\gamma_\psi s_i)\Phi 
\end{align}
for some Weyl tensor superfield $\Phi$ of weight $w=2$ \cite{Howe:1983fr, Koller:1982cs}. This is consistent with the dimension-2 Bianchi identity 
\begin{align}
	4\mathcal D_sH_{sss} - 4\cdot 3 i H_{ss\gamma(s,s)} = 0.
\end{align}
provided we also impose $H_{sss}=0$. 
Note that $H_{ss\psi}$ vanishes under the pure spinor projection. This is consistent with the limit of the dimension-$\tfrac52$ Bianchi identity
\begin{align}
	3\mathcal D_sH_{ss\psi} - \mathcal D_\psi H_{sss} +3\cdot 2 i H_{s\gamma(s,s)\psi} = 0,
\end{align}
from which we easily obtain
\begin{align}
	H_{s\psi\psi} = -(s^i\gamma_{\psi\psi} \mathcal D_i\Phi).
\end{align}
Therefore, in the notation of section \ref{S:Superforms}, $n=2$ and $n^\prime=1$. Plugging this into the dimension-3 identity 
\begin{align}
\label{E:3formBI3}
2\mathcal D_sH_{s\psi\psi} + 2\mathcal D_\psi H_{ss\psi} -2 i H_{\gamma(s,s)\psi\psi} + 4T_{s\psi}{}^{\alpha i} H_{\alpha i s\psi}= 0
\end{align}
and taking the pure spinor projection, we find the condition 
\begin{align}
\label{E:DDPhiTerms}
\tfrac18 (\lambda \gamma_{\psi\psi a} \lambda)\left( 
		\mathcal D^+\tilde \gamma^a \mathcal D^++ 16i C^{a++}
	\right) \Phi  = 0 .
\end{align}
Here, we have defined the combination 
\begin{align}
	\mathcal D^2_{a \, ij} :=\tfrac12 \mathcal D_{(i}\tilde \gamma^a \mathcal D_{j)}
\end{align} 
which will show up repeatedly. Thus, the dimension-3 identity implies the curved 3-form field strength constraint \cite{Linch:2012zh}
\begin{align}
\label{E:DDPhiConstraint}
		\mathcal D^2_{a\, ij}\Phi + 8i C_{aij} \Phi = 0
\end{align}
with the remaining combination defining the 3-form field strength component
\begin{align}
H_{\psi\psi\psi} = \tfrac i8 \left( 
	\mathcal D^k \tilde \gamma_{\psi\psi\psi} \mathcal D_k + 128i N_{\psi\psi\psi}
	\right) \Phi.
\end{align}

Acting on the constraint (\ref{E:DDPhiConstraint}) with $\mathcal D_s$ results in the Dirac equation
%\footnote
\begin{align}
\label{tensorWeyl}
i \mathcal D^{\alpha \beta} \mathcal D_{\beta}^i \Phi
	-  i C^{\alpha \beta\, ij} \mathcal D_{\beta j}\Phi
	-2i N^{\alpha \beta}\mathcal D_\beta^i \Phi
	-12i \mathcal C^{\alpha i}\Phi = 0
\end{align}
so this multiplet is on-shell.\footnote{An alternative interpretation of this formula in curved space is as a mechanism for defining geometrical objects in the Weyl multiplet in terms of those in the tensor multiplet \cite{Bergshoeff:1985mz} (see also \S2.3 of \cite{Linch:2012zh}).}
{The calligraphic torsion components entering here and below are the dimension-$\tfrac32$ components of the supergravity torsions (\ref{E:DCDN}). They are included here only for completeness and are not critical to the understanding of the 3-form.}
For completeness, we present the Klein-Gordon equation which results from contracting with another spinor derivative:
\begin{align}
\label{tensorKG}
0= \mathcal D^a\mathcal D_a \Phi 
	+8 C^a_{ij}C_a^{ij} \Phi 
	+\tfrac i6 N^{abc}\, \mathcal D^k\tilde \gamma_{abc} \mathcal D_k \Phi
	&-\tfrac{3i}2 \mathcal D_{\alpha i}\mathcal C^{\alpha i} \Phi  \cr
-3i \mathcal C^{\alpha i} \mathcal D_{\alpha i} \Phi
	&+  \tfrac{5i}2 \mathcal N^{\alpha i}\mathcal D_{\alpha i} \Phi.
\end{align}
The higher Bianchi identities
\begin{align}
\mathcal D_sH_{\psi\psi\psi} - 3\mathcal D_\psi H_{s\psi\psi} 
	- 3T_{s\psi}{}^{\alpha i} H_{\alpha i \psi\psi}
	- 3T_{\psi \psi}{}^{\alpha i} H_{\alpha i s \psi}= 0 \cr
4 \mathcal D_\psi H_{\psi\psi\psi} - \tfrac12\cdot 4\cdot 3 T_{\psi\psi}{}^{\alpha i} H_{\alpha i \psi\psi} =0	
\end{align}
do not imply any new conditions on $\Phi$ beyond those following from the constraint (\ref{E:DDPhiConstraint}). Instead of presenting calculations resulting in equations implied by (\ref{tensorWeyl}) and (\ref{tensorKG}), we will merely verify that the flat limits are identically satisfied. For the dimension-$\tfrac72$ identity:
\begin{align}
D_sH_{\psi\psi\psi} &=\tfrac12 (s^i \gamma^a \tilde \gamma_{\psi\psi\psi} D_i) \partial_a \Phi
	= \tfrac12 (s^i \gamma^a \tilde \gamma_{\psi\psi\psi} D_i
		- s^i \tilde \gamma_{\psi\psi\psi}\gamma^a  D_i) \partial_a \Phi\cr
	&=-3 \partial_\psi (s^i \gamma_{\psi\psi}D_i  \Phi )
		= 3 \partial_\psi H_{s\psi\psi} 
\end{align}
where we have used the Dirac equation on the spinor of $\Phi$. Similarly,
\begin{align}
\label{E:HClosure}
\partial_\psi H_{\psi\psi\psi} &=-\tfrac i{64}  \epsilon_{\psi\psi\psi\psi}{}^{ab} (D^i \tilde \gamma_{abc} D_i) \partial^c \Phi\cr
	&=-\tfrac i{64} \epsilon_{\psi\psi\psi\psi}{}^{ab} D^i (\tilde \gamma_{ab}\tilde \gamma_c -2\eta_{c[a} \tilde \gamma_{b]})D_i \partial^c \Phi \cr
	&=-\tfrac i{32} \epsilon_{\psi\psi\psi\psi}{}^{ab} D^i \tilde \gamma_{a}D_i \partial_b \Phi 
	=\tfrac 14 \epsilon_{\psi\psi\psi\psi}{}^{ab} \partial_a \partial_b \Phi \equiv 0,
\end{align}
by using the Dirac equation in the third equality.

%%%%%%%%%%%%%%%%%%%%%%%%%%%%%%%%
\subsubsection{Alternative formulation}
\label{S:3FormAlternative}
The recovery of the 3-form superfield as an obstruction to one of the two defining conditions  of the vector muliplet (\ref{E:PhiCondition}), suggests an alternative description of this form in terms of a dimension-$\tfrac32$ spinor superfield $V^{\alpha i}$ \cite{Bergshoeff:1996qm}. As we are obstructing only the scalar constraint, this field would still be required to satisfy the defining condition analogous to (\ref{E:XCondition1} or \ref{E:XCondition2}):
\begin{align}
\label{E:QV}
QV =0 
~~~\Leftrightarrow~~~
\mathcal D_{(i} \tilde \gamma_{ab} V_{j)} = 0.
\end{align}
It can be shown by brute force calculation that this condition is equivalent to the constraint (\ref{E:DDPhiConstraint}) with 
\begin{align}
\label{E:PhiDefinition}
\Phi = \mathcal D_{\alpha i} V^{\alpha i}.
\end{align}
This description neatly incorporates the gauge invariance of the 3-form field strength since $\Phi$ is invariant under $\delta V^{\alpha i} = \Lambda^{\alpha i}$ where $\Lambda$ satisfies both constraints (\ref{E:XCondition2}) and (\ref{E:PhiCondition}) defining the vector multiplet. The condition on $V$ implies that 
\begin{align}
\label{E:2FormComponents}
\mathcal D_{\alpha i} V^{\beta j} = \tfrac18 \delta_i^j \delta_\alpha^\beta \Phi 
	+ \tfrac14 \delta_i^j B_\alpha{}^\beta 
	+ \delta_\alpha^\beta Y_i{}^j
\end{align}
for a scalar $\Phi$, 2-form potential $B$, and an auxiliary triplet $Y$. Under the gauge transformation, the 2-form gauge field transforms into the field strength of a 1-form $\lambda$ as $\delta B = d\lambda$.

%%%%%%%%%%%%%%%%%%%%%%%%%%%%%%%%
\subsection{The closed $4$-form}
\label{S:4formFS}
The closure constraint (\ref{E:DDPhiTerms}) on the 3-form can be obstructed with a 4-form, the lowest non-trivial component of which is given by
\begin{align}
\label{E:4form}
G_{ss\psi\psi} = (s^i \gamma_{\psi\psi a} s^j) G^a_{ij}.
\end{align}
This is consistent with the dimension-$\tfrac72$ Bianchi identity 
\begin{align}
\label{E:4FormBI7/2}
	3\mathcal D_s G_{ss\psi\psi} 
	+ 3\cdot 2 i G_{s\gamma(s,s) \psi\psi} = 0
\end{align}
under the pure spinor projection, provided
\begin{align}
\label{E:GConstraint1}
	\Pi_{a\alpha}^{c\gamma} \, \mathcal D_{\gamma (k}G_{c\, ij)} = 0 
\end{align}
where $\Pi$ is the projector onto the $\gamma$-traceless subspace of the spinor-vector representation (\ref{E:Projector}).
To see this, note that the pure spinor projection of (\ref{E:4form}) is proportional to $(c \tilde \gamma^a c) G_a^{++}$. The condition that this be annihilated by $Q$ then becomes equivalent to the Weyl-invariant constraint (\ref{E:HigherSpin}) for $\ell=1$.
The remaining part of the Bianchi identity (\ref{E:4FormBI7/2}) is, then, easily solved %by {\it ansatz} 
for the dimension-$\tfrac72$ component
\begin{align}
	G_{s\psi\psi\psi} = - \tfrac i {12} \epsilon_{\psi\psi\psi}{}^{abc} (s^i \gamma_{ab} \mathcal D^j)G_{c ij} .
\end{align}

The dimension-4 Bianchi identity is 
\begin{align}
\label{E:4FormBI4}
2\mathcal D_s G_{s\psi\psi\psi} +3 \mathcal D_\psi G_{ss\psi\psi} 
	- 2 i G_{\gamma(s,s) \psi\psi\psi} 
	+ 2\cdot 3\, T_{s\psi}{}^{\alpha i} G_{\alpha i s \psi\psi} 
	= 0 .
\end{align}
The pure spinor projection gives three conditions:
\begin{align}
\label{E:GConstraint2}
\mathcal D^{2}_{a \,k(i}  G^a_{j)}{}^k + 24 i C_{a \,k(i}  G^a_{j)}{}^k +8i \mathcal D_a G^a_{\, ij} 
	=0,\\
\mathcal D^{2}_{(a \,k(i}  G_{b)\, j)}{}^k -4i \mathcal D_{(a} G_{b)\, ij} + \dots -\mathrm{trace} =0, \\
\mathcal D^{2}_{[a \,k(i}  G_{b]\, j)}{}^k -4i \mathcal D_{[a} G_{b]\, ij} 
	- \tfrac 18 \mathcal D^k\tilde \gamma_{abc}\mathcal D_k G^c_{ij} + \dots  =0,
\end{align}
where the ellipses stand for unilluminating torsion corrections.
The second and third condition follow from the constraint (\ref{E:GConstraint1}) by contraction with $(\mathcal D^k\tilde \gamma_b)^\alpha$. Indeed, the combination $\mathcal D^{2}_{a \,k(i}  G_{b\, j)}{}^k -4i \mathcal D_{a} G_{b\, ij} = -\tfrac 34 D^k\tilde \gamma_aD_{(k} G_{b\,ij)}$. Since $\Pi$ is $\gamma$-traceless, we get no condition upon contraction with $\eta^{ab}$. Therefore, the only condition not already implied by the lower Bianchi identities is the condition (\ref{E:GConstraint2}) on the trace. It is this condition that we will source to get the 5-form (c.f.~\S \ref{S:5formFS}).

The remaining terms in (\ref{E:4FormBI4}) determine the dimension-4 component
\begin{align}
\label{E:4formTop}
G_{\psi\psi\psi\psi} = 
	-\tfrac 1{24} \epsilon_{\psi\psi\psi\psi}{}^{ab} \left( \mathcal D^2_{a\, ij}  - 40 i C_{a\, ij}\right) G_{b}{}^{ij} .
\end{align}
In this calculation, there are no irreducible components beyond this 4-form that need to be canceled so we do not generate any additional constraints on $G_{a\,ij}$ at this level.

The remaining identities are the dimension-$\tfrac92$ identity 
\begin{align}
\mathcal D_s G_{\psi\psi\psi\psi} -4 \mathcal D_\psi G_{s\psi\psi\psi} 
	- 4 T_{s\psi}{}^{\alpha i} G_{\alpha i \psi \psi\psi} 
	-\tfrac12\cdot 4\cdot 3 T_{\psi\psi}{}^{\alpha i} G_{\alpha i s \psi\psi} 
	= 0 
\end{align}
and the dimension-5 identity
\begin{align}
5\mathcal D_\psi G_{\psi\psi\psi\psi} 
	-\tfrac12 \cdot 5\cdot 4 T_{\psi\psi}{}^{\alpha i} G_{\alpha i \psi \psi\psi} 
	= 0 .
\end{align}
They are satisfied identically in the flat limit. The closure of the top component implies that the dual 2-form $*G$ is divergenceless up to torsion terms. In the flat limit, $(* G)_{ab} =\tfrac 1{12} D^2_{[a}{}^{ij}  G_{b] \,ij}$ and it is straightforward to check that $\partial^b (* G)_{ab} \equiv 0$.

%%%%%%%%%%%%%%%%%%%%%%%%%%%%%%%%
\subsubsection{Alternative formulation} 
\label{S:4FormAlternative}
In section \ref{S:3FormAlternative} we explored the alternative ``potential'' formulation of the gauge 2-form. There, the condition defining the representation was expressed as $QV=0$ (\ref{E:QV}) instead of the condition (\ref{E:DDPhiConstraint}) in terms of its field strength $\Phi$. Similarly, one expects to be able to obstruct the closure condition in this potential-type formulation by taking
\begin{align}
\label{E:QVC}
QV = {C} .
\end{align}
Since $V = \bar \lambda_\alpha v_i V^{\alpha i}$ is a field with $(f,s,c) = (1,0,1)$ and $Q$ is an operator of type $(1,1,0)$, ${C}$ is of type $(2,1,1)$, that is, ${C}:= (\lambda \gamma^{ab} \bar \lambda)  v^iv^j \, {C}_{ab\, ij}$. Consistency then implies
\begin{align}
Q{C} = 0
~~\Rightarrow~~
w = 2
\end{align}
where the weight, again, follows from the general formula (\ref{E:Q=0}).

%\footnote{
Recall that the vector multiplet field strength $W$ obeys two conditions (\ref{E:XCondition1}) and (\ref{E:PhiCondition}). Relaxing (\ref{E:PhiCondition}) introduces the potential $V$ for the 3-form field strength which still obeys (\ref{E:QV}). We are now relaxing this second condition by introducing the potential $C$ for the 4-form field strength $G=dC$. The constraints on $G$ imply that $C_{ss\psi} = s^i\gamma_{\psi}{}^{ ab} s^j C_{ab\, ij}$ is the lowest non-vanishing component of this potential. The pure spinor projection of this component with $C_{ab\,ij} := \tfrac 18 D_{(i}\tilde \gamma_{ab} V_{j)}$ gives back equation (\ref{E:QVC}).

%%%%%%%%%%%%%%%%%%%%%%%%%%%%%%%%
\subsection{The closed $5$-form}
\label{S:5formFS}
The obstruction to closure of the 4-form is the left-hand side of (\ref{E:GConstraint2}). Our procedure, then, implies that the lowest component of the closed 5-form is given in terms of a superfield $K_{ij}$ by 
\begin{align}
K_{ss\psi\psi\psi} = s^i \gamma_{\psi\psi\psi} s^j K_{ij}. 
\end{align}
This is consistent with the dimension-$\tfrac92$ Bianchi identity 
\begin{align}
3\mathcal D_s K_{ss\psi\psi\psi} - 3\mathcal D_\psi K_{sss\psi\psi} + 3\cdot 2 i K_{s\gamma(s,s)\psi\psi\psi} =0
\end{align}
in the pure spinor projection provided 
\begin{align}
\label{E:5formConstraint}
QK = 0
~~~\Leftrightarrow~~~
\mathcal D_{\gamma(k} K_{ij)} = 0.
\end{align}
This condition is Weyl invariant when $w=4$ (\ref{E:WeylWeight}) in agreement with the engineering dimension of $K_{ss\psi\psi\psi}$. This analyticity constraint implies that
\begin{align}
\label{E:5formConstraint2}
\mathcal D^{2}_{a  (i}{}^k  K_{j) k}  +24i C_{a(i}{}^k K_{j)k} + 4i \mathcal D_a K_{ij}  =0
\end{align}
and
\begin{align}
\label{E:5formConstraint3}
\mathcal D^k \tilde \gamma_{abc} \mathcal D_k K_{ij} -128i N^{(-)}_{abc}K_{ij} = 0.
\end{align}
The remaining part of the Bianchi identity defines the dimension-$\tfrac92$ component of $K$ to be
\begin{align}
K_{s\psi\psi\psi\psi} = -\tfrac i{12} \epsilon_{\psi\psi\psi\psi}{}^{ ab}(s^i\gamma_{ab}\mathcal D^j) K_{ij} .
\end{align}

The dimension-5 Bianchi identity 
\begin{align}
2 \mathcal D_s K_{s\psi\psi\psi\psi} 
	+ 4 \mathcal D_\psi K_{ss\psi\psi\psi} 
	-2i K_{\gamma(s,s)\psi\psi\psi\psi} 
	-8 (T_{s\psi}{}^i\gamma_{\psi\psi\psi} s^j) K_{ij}= 0
\end{align}
is identically satisfied in the pure spinor limit due to the constraint (\ref{E:5formConstraint3}) and cancelation of the $\mathcal D_\psi$ terms. The remaining part determines the top component of $K$ to be
\begin{align}
\label{E:Ktop}
K_{\psi\psi\psi\psi\psi} = \tfrac1{24} \epsilon_{\psi\psi\psi\psi\psi}{}^{ a} \left(
	\mathcal D^2_{a\, ij} -48i C_{a ij} \right)K^{ij} .
\end{align}

The dimension-$\tfrac{11}2$ identity is
\begin{align}
\mathcal D_s K_{\psi\psi\psi\psi\psi} - 5\mathcal D_\psi K_{s\psi\psi\psi\psi}  
	- 5T_{s\psi}{}^{\alpha i} K_{\alpha i \psi\psi\psi\psi} 
	-\tfrac12 \cdot 5\cdot 4 T_{\psi\psi}{}^{\alpha i} K_{\alpha i s \psi\psi\psi} 
	=0.
\end{align}
It serves only to define the $\theta^3$-terms in $K$ in terms of space-time derivatives acting on the lower components and is otherwise unilluminating. 

The dimension-6 identity is
\begin{align}
6\mathcal D_\psi K_{\psi\psi\psi\psi\psi}  - \tfrac12\cdot 6\cdot 5 T_{\psi\psi}{}^{\alpha i} K_{\alpha i \psi \psi\psi\psi}  =0.
\end{align}
It tells us that, in the flat limit, the bosonic projection of the top component of the 5-form $K$ is closed in the bosonic sense. We may check this explicitly by using the flat covariant derivative identity $\partial^a D^{2}_{a\, ij} = \tfrac i{12}D^{3\alpha}_{ijk}\, D^k_\alpha$. It implies that the dual form $*K_a$ is divergenceless
\begin{align}
\partial^a (*K)_a = 0
\end{align}
due to the analyticity constraint (\ref{E:5formConstraint}) on $K$.

Another way to understand this result is by comparison with the 1-form of section \ref{S:1formFS}. 
In the flat limit, the constraint (\ref{E:5formConstraint3}) agrees with the defining condition (\ref{E:KollerSolution}) of a (gauge) 1-form prepotential. This implies that there is a 1-form at the $\theta^2$-level of $K$. Since $K$ is a field strength, and due to the dimension of this component, this vector must be a field strength. That this component is divergenceless where that of the vector multiplet was not is a consequence of the stronger constraint (\ref{E:5formConstraint}) (from which (\ref{E:5formConstraint3}) follows).

%%%%%%%%%%%%%%%%%%%%%%%%%%%%%%%%
\subsection{The closed $6$-form}
\label{S:6formFS}
The top component of the 5-form $K$ defined in (\ref{E:Ktop}) solves the dimension-5 Bianchi identity with no additional requirements on the superfield $K_{ij}$ beyond the defining pure spinor condition (\ref{E:5formConstraint}). As there is no obstruction to the closure of the 5-form, our procedure does not generate a non-vanishing 6-form at this level. 

We may nevertheless force the violation of the 5-form Bianchi identity by obstructing the defining relation and attempting to interpret the result as a closed 6-form. This corresponds to the {\it ansatz} 
\begin{align}
L_{sss\psi\psi\psi} = (s^i \gamma_{\psi\psi\psi}s^j)s^{\alpha k} L_{\alpha ijk}.
\end{align}
Upon pure spinor projection, this gives\footnote{Note the similarity of this expression with the pure spinor 0-mode normalization $\langle \lambda^3 \theta^5 \rangle =1$ \cite{oai:arXiv.org:hep-th/0001035}.}
\begin{align}
L_{sss\psi\psi\psi} \mapsto \lambda^\alpha (\lambda \gamma_{\psi\psi\psi}\lambda)  v^i v^j v^k L_{\alpha ijk}.
\end{align}
The projection of the lowest-dimension Bianchi identity (dimension 5)
\begin{align}
4\mathcal{D}_{s}L_{sss\psi\psi\psi} - 4\cdot 3i L_{ss\gamma(s,s)\psi\psi\psi} = 0, 
\end{align}
as usual, requires $QL=0$ which is again a condition of type (\ref{E:Q=0}). Explicitly,
\begin{align}
\label{E:6Form}
QL=0
	~~~\Leftrightarrow~~~ 
\mathcal D_{(\alpha (i} L_{\beta) jkl)} = 0 
	~~~\Rightarrow~~~ 
	w= \tfrac 92.
\end{align}
%%%%%%%%%%%%%%%
The remaining terms can be solved to find the dimension-5 component of the six form
\begin{align}
L_{ss\psi\psi\psi\psi} &= \tfrac{i}{48}\epsilon_{\psi\psi\psi\psi}{}^{ab}(s^{i}\gamma_{ab})^{\alpha}s^{\beta j} \left( 3\mathcal{D}_{\alpha }^{k}L_{\beta ijk} - \mathcal{D}_{\beta}^{k}L_{\alpha ijk}\right).
\end{align}

The dimension-$\frac{11}{2}$ Bianchi identity is
\begin{align}
3\mathcal{D}_{s}L_{ss\psi\psi\psi\psi} 
	- 4\mathcal{D}_{\psi}L_{sss\psi\psi\psi} 
	+3\cdot 2i L_{s\gamma(s,s)\psi\psi\psi\psi} 
	-3\cdot 4T_{s\psi}{}^{\alpha i}L_{\alpha i ss\psi\psi\psi}
	=  0.
\end{align}
The constraints following from the pure spinor projection at this, and at any other level in the Bianchi identities, %for this form 
can be obtained by hitting (\ref{E:6Form}) with derivatives.\footnote{This is because (\ref{E:6Form}) does not project out any irreducible component of $Q$ acting on $L$. Note that this is in contrast to (\ref{E:GConstraint1}) which projects out the $\gamma$-trace of $\mathcal D_{\gamma (k}G_{a \, ij)}$.}
We solve this Bianchi identity for the next component to find
\begin{align}
L_{s\psi\psi\psi\psi\psi} &= \tfrac{i}{192} \epsilon_{\psi\psi\psi\psi\psi}{}^{a}
      s^{\alpha i}\left(3\, \mathcal{D}^{2}_{a}{}^{jk}
        +124\,C_{a}{}^{jk} \right) L_{\alpha ijk}  \cr 
&- \tfrac{i}{192}\epsilon_{\psi\psi\psi\psi\psi}{}^{a} (s^i\gamma_{ab})^\alpha
   \left( \mathcal{D}^{2b\, jk} + 52\, C^{b\, jk} \right)L_{\alpha ijk}.
\end{align}
The dimension-6 Bianchi
\begin{align}
2\mathcal{D}_{s}L_{s\psi\psi\psi\psi\psi} 
	&+ 5\mathcal{D}_{\psi}L_{ss\psi\psi\psi\psi} 
	-2i L_{\gamma(s,s)\psi\psi\psi\psi\psi} \cr
	&+ 2\cdot 5T_{s\psi}{}^{\alpha i}L_{\alpha i s\psi\psi\psi\psi} 
	-\tfrac 12 \cdot 5\cdot 4 T_{\psi\psi}{}^{\alpha i}L_{\alpha iss\psi\psi\psi} 
	=0,
\end{align}
can be solved for the top component of the 6-form to give
\begin{align}
L_{\psi\psi\psi\psi\psi\psi} 
	&=\tfrac{i}{192} \epsilon_{\psi\psi\psi\psi\psi\psi}\left[ 
		\left( {\mathcal{D}}^{3 \alpha ijk} -12i\mathcal{C}^{\alpha ijk}\right)L_{\alpha ijk} 
		+10i C_a^{ij}(\mathcal{D}^{k}\tilde \gamma^aL_{ijk} )
		\right],
\end{align}
where $\mathcal D^{3\alpha}_{ijk}:= \tfrac1{4!}\varepsilon^{\alpha \beta \gamma \delta} \{ \mathcal D_{\delta(i}, [ \mathcal D_{\gamma j}, \mathcal D_{\beta k)}]\}$.
%{\color{red} Is this the definition you are using?}
The dimension-$\frac{13}{2}$ Bianchi
\begin{align}
\mathcal{D}_{s}L_{\psi\psi\psi\psi\psi\psi} - 6\mathcal{D}_{\psi}L_{s\psi\psi\psi\psi\psi} 
	- 6T_{s\psi}{}^{\alpha i}L_{\alpha i \psi\psi\psi\psi\psi} - \tfrac12\cdot 6\cdot 5T_{\psi\psi}{}^{\alpha i}L_{\alpha i s\psi\psi\psi\psi}=0.
\end{align}
%{\color{red}This looks wrong again!}\\
does not define any new components and serves only to define the $\theta^{4}$-terms in $L$ in terms of derivatives acting on its lower components. Similarly, the dimension-7 Bianchi
\begin{align}
7\mathcal{D}_{\psi}L_{\psi\psi\psi\psi\psi\psi} -\tfrac 12 \cdot 7 \cdot 6 T_{\psi\psi}{}^{\alpha i}L_{\alpha i \psi\psi\psi\psi\psi}=0
\end{align}
provides the bosonic closure condition (up to torsion) for the six form.
%{\color{red}Check this!}

We can solve the condition (\ref{E:6Form}) analogously to what was done in section \ref{S:1formFS} by taking 
\begin{align}
\label{Lprepot}
L_{\alpha ijk} = \mathcal D_{\alpha (i} L_{jk)} + \mathcal D_\alpha^l L_{ijkl}
\end{align}
and plugging it back in. Note that the first term is $Q$-exact and is therefore not constrained by (\ref{E:6Form}).  However, precisely analogously to the case of the constrained prepotential $U_{ij}$ of the gauge field (\ref{E:KollerSolution}), the field $L_{ijkl}$ must satisfy the condition 
\begin{align}
\mathcal D^m \tilde \gamma_{abc} \mathcal D_m L_{ijkl} -384i N^{(-)}_{abc}L_{ijkl} =0
\end{align}
for the constraint (\ref{E:6Form}) to hold.
In flat space, then, the analogue of Mezincescu's unconstrained prepotential $u^{ij}$ for the 6-form would be an unconstrained, dimension-2 scalar field $\ell$ such that $L_{ijkl} = D^4_{ijkl} \ell$.

%{\color{red}Check this!}

%%%%%%%%%%%%%%%%%%%%%%%%%%%%%%%%%%%%%%%%%%%%%%%%%%%%%%%%%%%%%%%%%%
%%%%%%%%%%%%%%%%%%%%%%%%%%%%%%%%%%%%%%%%%%%%%%%%%%%%%%%%%%%%%%%%%%
\section{Composite forms}
\label{S:Composite}
In the previous section, we constructed the de Rham  complex of differential forms by sequentially obstructing the closure condition with a form of degree 1 higher. In this section, we investigate the alternative method of building higher-degree forms by wedging forms of lower degree. 
Analogously to how solving the seemingly trivial closure conditions $d\omega=0$ resulted in the elucidation of the superspace representations of superconformally covariant $p$-form fields and their coupling to gravity, here we will similarly gain insight into the structure of interactions in superconformal $N=(1,0)$ models and their Lagrangians. In the process, we will derive relations between certain types of composite forms that we compare in section \ref{S:TensorHierarchy} to explicit formul\ae{} appearing in the non-abelian tensor hierarchy. 

We will refer to the forms obtained by wedging lower-degree forms as {\em composite} forms to distinguish them from the forms above. To minimize additional notation, we will use the same letters in bold font to denote the composite forms. 
Consider the the composite $p$-form $\bm\omega_p = \omega_q\wedge \omega_{p-q}$. For simplicity of exposition, we mostly focus on the product of only two forms. Then 
\begin{align}
\bm\omega_{s_1\dots s_s \psi_1\dots \psi_{p-s} }= \sum_{r+ t= s} c^s_{rt}\, \omega_{s_1\dots s_r\psi_1\dots \psi_{q-r} }\omega_{s_1\dots s_t\psi_1 \psi_{p-q-t}}
\end{align}
for some rational coefficients $c^s_{rt}$. These are computed by first counting inequivalent permutations of indices and then normalizing the result to 1.
%\footnote
{For example, the $\mathbf G_{ss\psi\psi}$ component of $\mathbf G=F\wedge F$ is gotten by writing down the terms $F_{ss}F_{\psi\psi}$ and $F_{s\psi}F_{s\psi}$ and realizing that there are two inequivalent configurations of the indices on the second term, namely $F_{\underline{\alpha} a}F_{\underline{\beta} b}$ and $F_{\underline{\beta} a}F_{\underline{\alpha} b}$ whereas on the first $F_{\underline{\alpha} \underline{\beta}}$ is equal to $F_{\underline{\beta}\underline{\alpha} }$ and similarly for $F_{ab}$. Therefore these 3 terms are weighted as $\tfrac13 F_{ss}F_{\psi\psi}$ and $\tfrac23 F_{s\psi}F_{s\psi}$. They get a relative sign from the odd permutation $ss\psi\psi\to s\psi s\psi$. Finally, $F_{ss}=0$ so $\mathbf G_{ss\psi\psi} = -\tfrac23 F_{s\psi}F_{s\psi}$.}

%%%%%%%%%%%%%%%%%%%%%%%%%%%%%%%%
\subsection{The composite $p$-form with $p=2$ and $3$}
In the abelian limit $\mathbf F := A\wedge A \equiv 0$ so that a single 1-form does not generate a composite 2-form. Given a collection of such forms and a bilinear, skew-symmetric map $\mathsf f$, however, one can construct $\mathbf F_{ss} := \mathsf f( A_s,A_s)$ and its higher components. If, in addition, $\mathsf f$ maps back into the collection of forms, we can use this composite 2-form as a deformation of the collection of abelian field strengths $dA$. If one further requires that these maps satisfy the Jacobi identity $\mathsf f ( \mathsf f(A_s, A_s), A_s) =0$ then this component can be absorbed into a connection $\nabla = D + \mathsf f (A, \cdot)$ and we recover the usual formulation of the non-abelian gauge field strength. 

The condition $F_{ss}=0$ is equivalent to $\nabla_s^2 = i \nabla_{\gamma(s,s)}$, defining the vector connection in terms of the spinor connection. With this, the first Bianchi identity becomes equivalent to the associativity of the spinor connection: $0= \nabla_s (\nabla_s \nabla_s) - (\nabla_s \nabla_s) \nabla_s = i [\nabla_s, \nabla_{\gamma(s,s)}] = i F_{s\gamma(s,s)}$. The rest of the analysis proceeds as in section (\ref{S:2formFS}).
  
A composite 3-form is easily constructed as $\mathbf H = A\wedge F$.\footnote{In the non-abelian case, this can be extended to the full Chern-Simons 3-form. For simplicity of exposition, we work in the abelian limit but allow $A$ and $F$ to be independent fields.} The properly normalized components
\begin{align}
\mathbf H_{sss} &= A_s F_{ss} = 0\cr
\mathbf H_{ss\psi} &= \tfrac 23 A_s F_{s\psi} + \tfrac13 A_\psi F_{ss} = \tfrac 23 A_s F_{s\psi}\cr
\mathbf H_{s\psi\psi} &= \tfrac 13 A_s F_{\psi\psi} - \tfrac23 A_\psi F_{s\psi} \cr
\mathbf H_{\psi\psi\psi} &= A_\psi F_{\psi\psi}
\end{align}
satisfy the Bianchi identities provided $A$ and $F$ satisfy theirs. That is, $\mathbf H$ is closed provided both $A$ and $F$ are. When $dF=0$ but $dA\neq0$, a short calculation 
\begin{align}
2(d\mathbf H)_{ss\psi\psi} &= 2 D_s \mathbf H_{s\psi\psi} + 2 \partial_\psi \mathbf H_{ss\psi} -2i \mathbf H_{\gamma(s,s)\psi\psi} \cr
	&= \tfrac 23 D_s A_s F_{\psi\psi} -\tfrac 23 A_s  D_s F_{\psi\psi} 
		- \tfrac43 D_s A_\psi F_{s\psi} - \tfrac43  A_\psi D_s F_{s\psi} \cr
	&+ \tfrac 43 \partial_\psi A_s F_{s\psi} +\tfrac 43  A_s \partial_\psi F_{s\psi} 
		-\tfrac{2i}3 A_{\gamma(s,s)} F_{\psi\psi} +\tfrac{4i}3 A_\psi F_{\gamma(s,s) \psi}  \cr
	&=  -\tfrac 23 A_s  \left ( D_s F_{\psi\psi} -2 \partial_\psi F_{s\psi} \right)
		- \tfrac43 \left( D_s A_\psi - \partial_\psi A_s  \right) F_{s\psi} \cr
	&- \tfrac23  A_\psi \left( D_s F_{s\psi} -2i F_{\gamma(s,s) \psi} \right)\cr
	&= -\tfrac43 (dA)_{s\psi} F_{s\psi}
\end{align}
shows that $(d\mathbf H)_{ss\psi\psi} = \mathbf G_{ss\psi\psi}$ where $\mathbf G = dA\wedge F$.

On the pure spinor subspace, this form is represented simply by the abelian Chern-Simons super-3-form field $\mathbf {C} = AW$. This is a composite analogue of the alternative description of the exact 4-form obstruction (\ref{E:QVC}). We will use this form in section \ref{S:TensorHierarchy} to obstruct the defining condition (\ref{E:QV}) of the gauge 2-form potential $V$ as
\begin{align}
	QV = \alpha\, \mathrm{tr}( AW)
\end{align}
for some parameter $\alpha$. This equation was proposed in flat space in the form $D_{(i} \tilde \gamma_{ab} V_{j)} = \alpha \, A_{(i} \tilde \gamma_{ab} W_{j)}$ in reference \cite{Bergshoeff:1996qm}. There, it was explained that this deformation is consistent since both sides obey the constraint $D_{(\gamma}^{(i} \sigma^{jk)}{}_{\beta)}^\alpha -\mathrm{traces} =0$ where $\sigma$ stands for the $DV$ and $AW$ combinations on the left-hand side and the right-hand side, respectively. In the pure spinor notation, this observation reduces to the fact that $Q^2=0$ on $V$ and that $Q(AW) = (QA) W + AQW =0$ by the defining equations (\ref{E:QA}) and (\ref{E:XCondition1}).

%%%%%%%%%%%%%%%%%%%%%%%%%%%%%%%%
\subsection{The composite $p$-form with $p=4,5,$ and $6$}
\label{S:Composite456}
Let ${{Z}}^{\alpha i}$ denote a positive chirality Weyl tensor of weight $\tfrac32$ and define its weight-2 field strength
\begin{align}
\bm \Phi({{Z}})  := \mathcal D_{\alpha i} {{Z}}^{\alpha i} .
\end{align}
Recall that when ${Z}$ satisfies the condition $Q{Z}=0$ (\ref{E:QV}), its associated field strength $\bm \Phi$ satisfies the condition (\ref{E:DDPhiConstraint}). Additionally, restricting $\bm \Phi=0$, implies that ${Z}$ describes the vector multiplet of section \ref{S:2formFS}.

For any two such spinor superfields $\tilde {Z}$ and ${Z}$, define the bilinear
\begin{align}
\label{E:Current}
\mathbf G_{a ij}(\tilde {{Z}}, {{Z}})  :=  \tilde {{Z}}_{(i} \gamma_a {{Z}}_{j)}.
\end{align}
When $\tilde {Z}, {Z}$ satisfy the condition (\ref{E:QV}), as we will henceforth assume, this bilinear satisfies (\ref{E:GConstraint1}) and  defines a composite version of the 4-form of section \ref{S:4formFS}. To see this, let $\mathbf B_{ab}(Z) := \mathcal D_i \tilde \gamma_{ab} Z^i$ denote the 2-form superfield associated to $Z$ in analogy to the definition of the fundamental 2-form (\ref{E:2FormComponents}). The composite 4-form has as its lowest non-trivial component $\mathbf G_{ss\psi\psi} = -\tfrac23 \tilde {\mathbf B}_{s\psi} \mathbf B_{s\psi} $. In the pure spinor projection this component becomes proportional to $(\lambda \gamma_{\psi\psi a} \lambda)v^i v^j \mathbf G^a_{ij}$.

In section \ref{S:4formFS} we found that the top component of the 4-form is given by (\ref{E:4formTop}). Consider the composite version in the flat limit
\begin{align}
(*\mathbf G)_{ab} := D_{[a ij} \mathbf G_{b]}{}^{ij} ,
\end{align}
the bosonic part of which evaluates to
\begin{align}
\label{E:GBose}
\mathbf G|_\mathrm{bose}  =
	-\tfrac3{16} \left[ 
		\tilde {\bm \Phi} * {\mathbf B} + \bm \Phi*\tilde  {\mathbf B}
		+6\tilde  {\mathbf  B}\wedge  {\mathbf  B}
		\right] .
\end{align}
In general, this form is not closed. Indeed, straightforward $D$-algebra gives
\begin{align}
\partial^b (*\mathbf G)_{ab}
	=\tfrac 18 D_{a ij} \left( \mathbf K^{ij}  + \tilde {\mathbf K}^{ij} \right)
\end{align}
where we have defined the bilinear $\mathbf K_{ij} := i D_{\alpha (i}\tilde {\bm \Phi}  {Z}^\alpha_{j)}+\tfrac i4 \tilde {\bm \Phi} D_{\alpha (i} {Z}^\alpha_{j)}$ and the $\tilde {\mathbf K}$ that follows from switching $\tilde Z \leftrightarrow Z$. This combination, or its curved version 
\begin{align}
\label{E:LComposite}
	\mathbf K_{ij} = i \mathcal D_{\alpha (i}\tilde {\bm \Phi}  {Z}^\alpha_{j)}+ \tfrac i4 \tilde {\bm \Phi} \mathcal D_{\alpha (i} {Z}^\alpha_{j)},
\end{align}
is a composite analogue of the defining field of the 5-form multiplet of section \ref{S:5formFS}. It is analytic ({\it i.e.} it satisfies equation \ref{E:5formConstraint}) because $\tilde {\bm \Phi}$ satisfies (\ref{E:DDPhiConstraint}) \cite{Linch:2012zh}. 
Since $\mathbf K(\tilde Z, Z) \neq \mathbf K(Z,\tilde Z)$ is not symmetric as a function of $\tilde Z$ and $Z$, the divergenceless vector superfield
\begin{align}
D^2_{a ij} \mathbf K^{ij} = 
	\tfrac 32\tilde \Phi {\stackrel\leftrightarrow \partial}_a \Phi 
	-\tfrac32 \partial^b \left( \tilde \Phi B_{ab}\right) 
	-\tfrac34 \tilde H^{(-)}_{abc} B^{bc} 
	+\mathrm{fermions}.
\end{align}
gives rise to {\em two} conserved currents when there are at least two tensor fields present. This will be important when we interpret our complex in terms of the non-abelian tensor hierarchy in section \ref{S:TensorHierarchy}.

The composite 4-form superfield (\ref{E:Current}) can be used to obstruct the defining constraint on the fundamental ({\it i.e.} not composite) 3-form field strength superfield $\Phi$:
\begin{align}
\label{E:DDPhiDeformed}
		\mathcal D_{(i}\tilde \gamma_a \mathcal D_{j)}\Phi + 16i C_{a ij} \Phi = \alpha \mathbf G_{a ij}.
\end{align}
where $\alpha$ is a coupling constant. We now turn to the analysis of this deformation in the case where $\tilde Z$ and $Z$ are some combination of vector and tensor multiplets.

%%%%%%%%%%%%%%%%%%%%%%%%%%%%%%%%
\subsubsection{The composite 4-form}
\label{S:4formComposite}
Specializing $\tilde {Z}={Z} =W$ to a single vector multiplet, $\mathbf G_{a ij}$ becomes the usual supercurrent \cite{Howe:1981qj, Howe:1983fr}.\footnote{In six dimensions, this current is analytic only on-shell as $\mathcal D_{\gamma (k} \mathbf G^a_{ij)}$ is proportional to the derivative of the vector multiplet auxiliary field $X_{ij}$.}
In this special case, the composite 4-form (\ref{E:GBose}) reduces to $\mathbf G \sim F\wedge F$. It is closed (the 4-form Bianchi identities are not obstructed) since $\bm \Phi\equiv 0$ and, therefore also, $\mathbf K\equiv 0$. The construction is off-shell as there are no tensors present to put it on-shell.

If we couple this form to a fundamental tensor, we recover the fact that the obstructed closure condition (\ref{E:DDPhiDeformed}) is the superspace analogue of the Green-Schwarz anomaly equation \cite{Bergshoeff:1996qm}
\begin{align}
\label{E:GS}
dH = \alpha F\wedge F.
\end{align}

%%%%%%%%%%%%%%%%%%%%%%%%%%%%%%%%
\subsubsection{The composite $5$-form}
In this section, we take $\tilde {Z}=V$ and $ {Z}=W$ to describe a tensor multiplet and a vector multiplet, respectively. The composite $\mathbf G_{a ij}$ still describes a 4-form (\ref{E:GBose}) but now in terms of a gauge 2-form $\tilde {\mathbf B} \to B$ and a field strength 2-form ${\mathbf B} \to F$. The associated composite field strength $\mathbf K$ is sourced by the analytic vector-tensor multiplet Lagrangian $-i\mathbf K_{ij} = \Phi X_{ij} + \mathcal D_{(i} \Phi W_{j)}$ (\ref{E:LComposite}).

This linear multiplet has an interpretation as a composite version of the super-5-form: Consider the lowest component of the composite form $\mathbf  K=F\wedge H$,
\begin{align}
\mathbf K_{ss \psi\psi\psi} 
	&=\tfrac 1{10}F_{ss} H_{\psi\psi\psi} -\tfrac6{10} F_{s\psi} H_{s \psi\psi} + \tfrac 3{10}F_{\psi\psi} H_{ss \psi} \cr
	&=-\tfrac35 F_{s\psi} H_{s \psi\psi} + \tfrac 3{10}F_{\psi\psi} H_{ss \psi}.
\end{align} 
In the pure spinor projection, this reduces to
\begin{align}
	\mathbf K_{\lambda\lambda \psi\psi\psi} = \tfrac{2i}5 \lambda \gamma_{\psi\psi\psi}\lambda\, v^iv^j\,\mathbf K_{ij}+Q\textrm{-exact},
\end{align}
that is, the pure spinor projection of the lowest component of the super-5-form is proportional to $\mathbf K_{ij}$ up to a $Q$-exact term.\footnote{This is not surprising since $Q\mathbf  K_{\lambda \lambda \psi\psi\psi}\propto Q F_{\lambda \psi}H_{\lambda \psi\psi} + F_{\lambda \psi}QH_{\lambda \psi\psi}=0$ by Bianchi identities and the combination $\mathbf K_{ij}$ was originally constructed in reference \cite{Linch:2012zh} to satisfy specifically this condition.}%end footnote

The top component of the 5-form $\mathbf K_{\psi\psi\psi\psi\psi}$ is given in terms of $\mathbf K_{ij}$ in equation (\ref{E:Ktop}). Here, we will explicitly compute its flat-space dual $*\mathbf K_a = \tfrac1{48} D_{a ij} \mathbf K^{ij}$. 
The bosonic part evaluates to
\begin{align}
\label{E:Kbose}
\mathbf K_a \Big|_\mathrm{bose} 
	&= -\tfrac 18 \partial^b\left( \Phi F_{ab}\right) -\tfrac 1{16} H^{(-)}_{abc} F^{bc}
\end{align}
where we used the off-shell version of the Maxwell equation
\begin{align}
D_{a ij} X^{ij} = 6i\partial^b F_{ab}.
\end{align}
For the fermionic part, we need the equation
\begin{align}
D_{\alpha i} F_\beta{}^\gamma = 2i\partial_{\alpha \beta} W^\gamma_i +\tfrac 23 \delta_\alpha^\gamma D_\beta^j X_{ij} - \tfrac 13\delta_\beta^\gamma D_\alpha^j X_{ij}
\end{align}
and its consequence
\begin{align}
D_{a ij} W^{\gamma j} = 6i \partial_a W^\gamma_i - (\tilde \gamma_a)^{\gamma \delta}D_{\delta j} X_i{}^j .
\end{align}
With this,
\begin{align}
\label{E:Kfermi}
\mathbf K_a \Big|_\mathrm{fermi} =\tfrac 18 D_k \Phi {\stackrel{\leftrightarrow}{\partial}}_a W^k 
		+\tfrac 18 D_k\Phi \tilde \gamma_a\gamma_b\partial^b W^k 
\end{align}
where we have used the fact that the tensor multiplet is on shell (\ref{tensorWeyl}). It is also due to this condition that $\mathbf K$ is divergenceless: Algebraically,
\begin{align}
\partial^a \mathbf K_a =- \tfrac 1{16} \partial^a H^{(-)}_{abc} F^{bc}
	-\tfrac 18  D_k \Box\Phi  W^k 
	+\tfrac 18 \partial^a D_k\Phi \tilde \gamma_a\gamma_b\partial^b W^k
\nonumber	
\end{align}
and these terms are all proportional to the equations of motion (\ref{tensorWeyl}, \ref{tensorKG}, \ref{E:HClosure}) of the tensor multiplet.

%%%%%%%%%%%%%%%%%%%%%%%%%%%%%%%%
\subsubsection{The composite $6$-form}
\label{S:6formComposite}
Finally, we consider the case where both $\tilde {Z}$ and ${Z}$ describe tensor multiplets $\tilde V$ and $V$.
The composite 5-form density resulting from this double-tensor can be understood along the lines of the vector-tensor construction of the previous section by replacing $F\to \tilde B$. In particular, $\mathbf K = \tilde B\wedge H$ and the current (\ref{E:Kbose}, \ref{E:Kfermi}) gets modified to the form
\begin{align}
\label{E:BBform}
i D_{a ij} \mathbf K^{ij} &\sim  
	i\tilde \Phi {\stackrel \leftrightarrow \partial}_a \Phi
	+\partial^b\left( \Phi  \tilde B_{ab}\right) 
	+3i H^{(-)}_{abc} \tilde B^{bc} \cr
&+D^k\Phi \tilde \gamma_a D_k \tilde \Phi
	+D_k \Phi {\stackrel{\leftrightarrow}{\partial}}_a V^k 
	+D_k\Phi \tilde \gamma_a\gamma_b\partial^b V^k  .
\end{align}
Note that this composite is not gauge invariant. It also does not generate a gauge invariant 6-form since the current is conserved.
Conservation uses the equations of motion of both multiplets and the fact that $H\wedge \tilde H \equiv 0$ for any two anti-self-dual forms $H$ and $\tilde H$. In the case $\tilde \Phi = \Phi$, the first term on each line vanishes. We will return to this form in section \ref{S:TensorHierarchy}.

The fact that we do not generate a closed 6-form with this bilinear is the composite analogue of the observation in section \ref{S:6formFS} that there is no obstruction to the closure of the 5-form $K_{ij}$ once it satisfies the defining relation $QK=0$ (\ref{E:5formConstraint}). Similarly to the analysis of that section, we can nevertheless define such a composite provided we go beyond bilinears and construct the analogue of $F\wedge F\wedge F$: 
\begin{align}
\mathbf L_{sss\psi\psi\psi} = 
	\tfrac{3\cdot 3}{15} F_{ss}F_{s\psi}F_{\psi\psi}
	-\tfrac{3!}{15}F_{s\psi}F_{s\psi}F_{s\psi}
	=-\tfrac 25F_{s\psi}F_{s\psi}F_{s\psi}
\end{align}
In the pure spinor projection, this becomes 
\begin{align}
\mathbf L_{sss\psi\psi\psi} \mapsto  
	\tfrac{16i}5 \, c^{3 \alpha} v^iv^jv^k (W^3)_{\alpha ijk}
\end{align}
giving the composite analogue $\mathbf L_{\alpha ijk}$ of the closed 6-form field strength of section \ref{S:6formFS}. Note that the Weyl weight of this composite is $w = 3\cdot \tfrac32=\tfrac92$ in agreement with the condition (\ref{E:6Form}).

%%%%%%%%%%%%%%%%%%%%%%%%%%%%%%%%%%%%%%%%%%%%%%%%%%%%%%%%%%%%%%%%%%
%%%%%%%%%%%%%%%%%%%%%%%%%%%%%%%%%%%%%%%%%%%%%%%%%%%%%%%%%%%%%%%%%%
\section{Applications}
\label{S:Applications}

In the previous sections we studied the structure of differential forms in six-dimensional, $N=(1,0)$ superspace. In this section, we present a selection of applications of these results. Topics we have refrained from discussing include applications to 
covariant superstring compactifications \cite{oai:arXiv.org:1109.3200} and related superspace gauge theories ({\it e.g.} ref. \cite{Cederwall:2008zv}),
the new ambi-twistor strings of \cite{Mason:2013sva},
the construction of superconformal theories with a second, non-linearly realized supersymmetry, 
superspaces with boundaries \cite{Howe:2011tm}, 
and the comparison to interesting recent lower-dimensional results ({\it e.g.} ref. \cite{Kuzenko:2013rna, Butter}).
Instead, we restrict our attention to the two applications that most overlap with the results already derived.
These sections are intended only to motivate the use of superforms and do not represent complete analyses which are still in progress.

%%%%%%%%%%%%%%
\subsection{Ectoplasm}
\label{S:Ectoplasm}
This work has its origin in failed attempts to construct the density projection formula for curved, six-dimensional, $N=(1,0)$ projective superspace action \cite{EctoRef}
\begin{align}
\label{E:5Ecto}
S&= \frac1{2\pi} \oint_C (v_idv^i) \int_M d^6x \int d^8\theta E \, \Theta^{(-4)} \mathcal L^{++} .
\end{align}
Generally, the Bianchi identities are solved in curved superspace for a $p$-form with $p$ equal to the dimension of the bosonic space-time. By an extension of Noether's argument defining conserved charges from conserved currents, the components of this form can be shown to define a curved supersymmetric invariant extending the flat-space component action \cite{Siegel:1999ew}. This invariant is, then, a natural candidate for the component action in curved superspace.

Six-dimensional, $N=(1,0)$ superspace has the peculiar property of disallowing the straightforward construction of a natural 6-form. The na\"ive generalization of two known approaches immediately fails for trivial reasons. One of these extends the observation that it is sometimes possible to construct the top form by wedging two middle-dimensional forms \cite{Gates:2009xt}. Applied to six dimensions, we expect to obtain the top form corresponding to the projective measure defined in \cite{Linch:2012zh} from the wedge of the 3-form with itself. This fails, however, since the 3-form is self-dual so that the associated 6-form vanishes identically in the flat limit. In curved superspace, it fails to produce the $\mathcal D^4$ part of the analytic measure. An attempt to construct a 6-form from other composites ({\it e.g.} three 2-forms) does not generate a forth-order operator acting on a scalar Lagrangian and, therefore, also does not represent the curved analytic measure.

A second attempt to guess the 6-form directly may be made by using Berkovits' {\it ansatz} for the structure of the lowest component of the top form \cite{Berkovits:2006ik}. The proposed component is of the form $L_{\underline \alpha \underline \beta \underline \gamma abc} \sim (\gamma_{abc})_{(\underline \alpha \underline \beta} f_{\underline \gamma)}$ with $D_{(\underline \alpha} f_{\underline \beta)} =0$. However, this component is pure gauge when interpreted as a Weyl tensor superfield as described in section \ref{S:1formFS}. In flat superspace, the Berkovits conjecture can be modified in the Biswas-Siegel approach to $p$-forms in harmonic superspace \cite{Biswas:2001wu} by constructing a 7-form with one leg in the harmonic sphere $\mathbb CP^1$: $\mathcal L_{\pp \, \alpha-\beta-\gamma-\, abc} = (\gamma_{abc})_{(\alpha \beta} D_{\gamma)}^- {\mathcal L}^{++}$. The superfield ${\mathcal L}^{++}$ is required to be analytic $D^+_\alpha {\mathcal L}^{++}$ and the top component $*{K} \propto D^{-4} {\mathcal L}^{++}$ reproduces the flat limit of the projective measure of \cite{Linch:2012zh}. Somewhat surprisingly, however, the curved superspace Bianchi identities cannot be satisfied for this choice of 6-form: The non-trivial isospin structure of ${\mathcal L}^{++}$ forces the dimension-1 torsions to vanish. (For example, already the first Bianchi identity implies $0=\mathcal D^+_{(\alpha}\mathcal D_{\beta)}^- {\mathcal L}^{++} \propto (\gamma^{abc})_{\alpha \beta} N_{abc} \mathcal L^{++}$.)

While it is beyond the scope of this work to investigate the question of integration in projective/harmonic superspace in any depth, the application of our results on differential forms already suggests some preliminary insights. 
For example, the analysis of section \ref{S:5formFS} suggests that the action for a linear multiplet in curved superspace reduces to the component result
\begin{align}
\label{E:5Ecto}
\int_{N} \mathrm d^5x\, e\,n^a \left\{
		 \left( \mathcal D^2_{a\, ij} -48i C_{a\, ij}\right) K^{ij}
		+\tfrac{4i}5  \Psi^b_i \tilde \gamma_{ab} \mathcal D_j K^{ij}
		+\tfrac65\Psi^b_i \tilde \gamma_{abc} \Psi^c_j K^{ij} 
	\right\}
\end{align}
with the integral taken over some 5-dimensional bosonic subspace $N$ of $M$. The full analysis of the four-dimensional analogue of this was carried out in reference \cite{Butter:2012ze}.

In section \ref{S:6formFS}, we found that the correct {\it ansatz} for the 6-form was $L_{sss \psi\psi\psi} = s^i \gamma_{\psi\psi\psi} s^j s^{\gamma k} L_{\alpha ijk}$.
Together with the other components derived in that section, we can write down a supersymmetric invariant that, schematically, is given by 
\begin{align}
\label{E:6Ecto}
\int \mathrm d^6x \, e \left\{
		\mathcal D^{3} L
		+\Psi \mathcal D^{2} L 
		+\Psi \Psi \mathcal D L  
		+\Psi \Psi \Psi  L 
	\right\} .
\end{align}
If one further solves the constraint (\ref{E:6Form}) on the dimension-$\tfrac92$ component as in (\ref{Lprepot}), one obtains a formula for covariantizing the component $\mathcal D^4_{ijkl}L^{ijkl}$. The method used in \cite{{Kuzenko:2007hu}, {Kuzenko:2008ry}} to obtain the analogous density projection formula starts with precisely such a term and successively constructs the higher components in the gravitino expansion in a Noether-type procedure based on the invariance under projective $SL_2(\mathbb C)$ transformations of the projective superspace action (\ref{E:5Ecto}). Therefore, if the component result from ectoplasm can be checked to be $SL_2(\mathbb C)$-invariant, it should correspond to the density projection formula for the projective superspace action.

%%%%%%%%%%%%%%%%%%%%%%%%%%%%%%%%
\subsection{Abelian tensor hierarchy}
\label{S:TensorHierarchy}

The non-abelian tensor hierarchy \cite{Samtleben:2011fj, Samtleben:2012mi} is an attempt to construct a non-abelian gauge theory of forms of degree $p>1$ by obstructing the closure of the standard Yang-Mills field strength. As we review in appendix \ref{S:NonAbTensorHierarchy}, one introduces a collection of $p$-form potentials $(B^I, C_r, D_\alpha, E_\mu)$ for $p =2, 3,4, 5$, respectively, extending the standard Yang-Mills potential $A^r$. A collection of linear maps $(\mathsf h^r_I, \mathsf g^{rI}, \mathsf k^\alpha_r)$ is introduced to obstruct the closure of the $p$-form field strength with a $(p+1)$-form potential. Consistency of this deformation in the non-abelian case requires the extension of the Yang-Mills structure constants $\mathsf f_{rs}^t$ by a collection of constants denoted by $(\mathsf d^I_{rs}, \mathsf b_{Irs}, \mathsf c_{\alpha IJ}, \mathsf c^{\prime s}_{\alpha r})$. Finally, superpartners are introduced and the whole model is shown to be superconformally invariant. 

The first step in this program is the obstruction of the Bianchi identity of a $p$-form field strength with a $(p+1)$-form field strength. This is precisely the program carried out in section \ref{S:deRham} to derive the complex of differential forms. Thus, the linearized part of the non-abelian tensor hierarchy is just the construction of this complex. Consider, for example, the case of the vector multiplet field strength $W$. Shifting $W\to W+ \mathsf h(V)$ obstructs the Bianchi identity (\ref{E:PhiCondition}) by the term (\ref{E:PhiDefinition}). As explained in section \ref{S:3FormAlternative}, this is the superfield defining the 3-form field strength of section \ref{S:3formFS}. It satisfies the condition (\ref{E:DDPhiConstraint}) which can, in turn, be obstructed by $\mathsf g(G)$ using the 4-form field of section \ref{S:4formFS}. 

This interpretation of the differential complex refers only to the linearized part of the hierarchy. We see from equation (\ref{E:DDPhiDeformed}), however, that certain non-linear parts are captured by introducing the composite deformations from section \ref{S:Composite} alongside the fundamental ones. Indeed, if, after constructing the non-abelian tensor hierarchy, one takes the abelian limit $\mathsf f\to 0$, one is {\em apparently} left with a non-linear theory.\footnote{There is a subtlety concerning the non-triviality of this limit that we address in footnote \ref{F:Abelianization}.} %end footnote
It seems to be the case, then, that the non-linear but abelian part of the hierarchy is precisely the entire differential complex augmented with composite obstructions. In this sense, one may think of the non-abelian tensor hierarchy as a non-abelian deformation of this complex or ``non-abelian ectoplasm" in the terminology of section \ref{S:Ectoplasm}.

Although demonstration of the complete equivalence of the two sides and the non-abelian extension of them is beyond the scope of this paper, some non-trivial comparisons can be made with the results already worked out.
The obstructed closure condition (\ref{E:GS}) is central to the construction of the non-abelian tensor hierarchy in which it appears in the form
\begin{align}
\label{E:TH1}
	d\mathcal H =  {\sf d} (\mathcal F, \mathcal F) + \mathsf g (\mathcal G ).
\end{align}
Here, 
\begin{itemize}
\item $\sf d$ is the symmetric bi-linear form (extended to act by wedge product on forms) on the space of vector multiplets valued in the space of tensor multiplets, 
\item $\mathcal F = F + \mathsf h (B)$ is a deformation of the non-abelian 2-form field strength $F=dA$ by a gauge 2-form $B$,
\item $\mathcal H = dB + \mathsf g(C)$, is a deformation of the 3-form field strength $H=dB$ by a gauge 3-form $C$, and
\item $\mathcal G = dC + \mathsf k(D)$, is a deformation of the 4-form field strength $G=dC$ by a gauge 4-form $D$ although this term does not enter into the hierarchy at this level since $\mathsf g  \circ \mathsf k= 0$.
\end{itemize}
As we have seen in section \ref{S:4formFS} and \ref{S:4formComposite}, this condition results from sourcing the defining equation of the tensor superfield strength (\ref{E:DDPhiDeformed}). Therefore, provided we shift the pure spinor superfields $W\to \mathcal W= W + \mathsf h (V)$, we can capture the $\sf b, c, c^\prime =0$ sector of the hierarchy in curved superspace with the constraint
\begin{align}
\label{E:DDJanktiPhi}
\left(\mathcal D_{(i}\tilde \gamma_a \mathcal D_{j)} + 16i C_{a ij} \right)\Phi = {\sf d} (\mathcal W_{(i}\gamma_a \mathcal W_{j)}) + \mathsf g( G_{aij} ) 
\end{align}
defining the deformed 3-form $\mathcal H$.

It was shown in reference \cite{Bandos:2013sia} that the $p$-form field strengths with $p\geq 4$ are all {\em composite}. Therefore, at least when formulated in terms of field strengths, it is possible that this constraint already encodes the entire abelian hierarchy. In fact, the composite ``current'' (\ref{E:Current}) has a natural extension by the associative $*$-product of appendix \ref{S:NonAbTensorHierarchy} to $\mathbf G_{a ij}\sim \mathsf b( W, V) + \mathsf {k\circ c} ( V, V) $. Associated to this field is a composite linear superfield $\mathbf K_{ij}$ (\ref{E:LComposite}). 
An important set of constraints (compare eq. (3.7) of ref. \cite{Samtleben:2011fj}) in the non-abelian tensor hierarchy is given in superspace by setting
\begin{align}
	\mathsf b(\mathbf K_{ij}) =0.
\end{align}	
In the context of section \ref{S:Composite456}, this equation implies that the 4-form Bianchi identities are satisfied when the composite obstruction is mapped to the space of 4-forms by $\mathsf b$. 
Equations (\ref{E:Kbose}) and (\ref{E:Kfermi}) then imply the deformed closure condition (compare eq. (3.39) and (3.43) of reference \cite{Bandos:2013sia})
\begin{align}
d \mathbf G  = {\sf b} (F, H) + {\sf k}(\mathbf K),
\end{align}
where $\mathbf K$ stands for the terms given in equation (\ref{E:BBform}). These terms make up the composite 5-form of the tensor hierarchy (compare eq. (3.43) and (3.49) of reference \cite{Bandos:2013sia}) in the $c^\prime\to 0$ limit.\footnote{As we have restricted ourselves to quadratic bilinears in section \ref{S:Composite456}, we will not generate the $\mathsf c^\prime$ contributions of the full tensor hierarchy here.}
Thus, we have found that this level of the hierarchy is compactly described by equation (\ref{E:DDJanktiPhi}).

By (a deformation of) the discussion in section \ref{S:3FormAlternative}, the condition (\ref{E:GS}) is equivalent to 
\begin{align}
	QV = {\sf d}( A , W ),
\end{align}
provided we describe the tensor $\Phi$ in terms of its potential $V$ \cite{Bergshoeff:1996qm}. Using our shifted fields, we can attempt to write the analogous expression for (\ref{E:DDJanktiPhi}) in pure-spinor superspace. The na\"ive guess is $QV = \mathsf d( A , W ) +\mathsf  g(C)$ for the 3-form potential $C$ described in section \ref{S:4FormAlternative}. However, according to \cite{Bandos:2013sia}, the associated field strength $G\sim dC$ is {\em composite}. In fact, it is precisely the composite appearing because the Bianchi identities of dimension $\ge 3$ do not close. 
Therefore, it may even be that an equation of the form
\begin{align}
D \mathsf A + \mathsf A * \mathsf A  = 0
~~~\mathrm{with}~~~
D = Q + \partial
~~~\mathrm{and}~~~
\mathsf A \in \Omega^\bullet \otimes K_\bullet
\end{align}
by itself already describes the entire hierarchy in the abelian limit.\footnote{It may be of interest to recall here that the complex of $p$-forms was determined uniquely by applying the obstruction procedure to the condition $F_{ss}=0$. At the beginning of section \ref{S:2formFS}, it was explained that this condition is equivalent to the formula $QA = 0$.} Work is currently underway to confirm this statement and extend it to the full non-abelian hierarchy. 

%%%%%%%%%%%%%%%%%%%%%%%%%%%%%%%%%%%%%%%%%%%%%%%%%%%%%%%%%%%%%%%%%%
%%%%%%%%%%%%%%%%%%%%%%%%%%%%%%%%%%%%%%%%%%%%%%%%%%%%%%%%%%%%%%%%%%
\section{Acknowledgements}
It is a pleasure to thank 
Igor Bandos for discussions of his work,
Brenno Carlini Vallilo for discussions and collaboration relating this work to the ambi-twistor string of reference \cite{Mason:2013sva},
and Jim Gates for encouragement, support, and references.
We are especially indebted to 
Gabriele Tartaglino-Mazzucchelli for vetting this manuscript, his emphasis of subtle points we had not appreciated regarding integration in curved superspace, and his help with references
and Robert Wimmer for carefully reading a previous version of this work, detailed discussions clarifying aspects of the non-abelian tensor hierarchy, and many suggestions for improving the presentation.

This work was partially supported by the National Science Foundation grants PHY-0652983 and PHY-0354401 and F{\sc ondecyt} (Chile) grant number 11100425. 
C{\sc a} is supported by the U{\sc nab-dcf} M.Sc. scholarship.
W{\sc dl}3 is partially supported by the U{\sc mcp} Center for String \& Particle Theory. 
A{\sc kr} acknowledges participation in the 2013 Student Summer Theoretical Physics Research Session.  

%\newpage
%%%%%%%%%%%%%%%%%%%%%%%%%%%%%%%%%%%%%%%%%%%%%%%%%%%%%%%%%%%%%%%%%%
%%%%%%%%%%%%%%%%%%%%%%%%%%%%%%%%%%%%%%%%%%%%%%%%%%%%%%%%%%%%%%%%%%
%%%%%%%%%%%%%%%%%%%%%%%%%%%%%%%%%%%%%%%%%%%%%%%%%%%%%%%%%%%%%%%%%%
%%%%%%%%%%%%%%%%%%%%%%%%%%%%%%%%%%%%%%%%%%%%%%%%%%%%%%%%%%%%%%%%%%
\appendix

%%%%%%%%%%%%%%%%%%%%%%%%%%%%%%%%%%%%%%%%%%%%%%%%%%%%%%%%%%%%%%%%%%
%%%%%%%%%%%%%%%%%%%%%%%%%%%%%%%%%%%%%%%%%%%%%%%%%%%%%%%%%%%%%%%%%%
\section{Curved six-dimensional superspace}
\label{S:6DGeom}
In this appendix, we collect the results on six-dimensional, $N=(1,0)$ supergravity used in our analysis of superforms in curved space-time. A detailed understanding of this material is not absolutely necessary to follow the discussion in the main text and serves mainly to fix some notation and introduce the supergravity torsion fields. For additional details, see references \cite{Linch:2012zh} and \cite{CesarThesis}.

We denote the local coordinates on curved, six-dimensional, $N=(1,0)$ superspace by $(z^{M})=(\theta^{\mu i}, x^{m})$. The covariant derivative $(\mathcal D_{A})= (\mathcal D_{\alpha i}, \mathcal D_{a})$ expands out to
\begin{align}
\mathcal D_{A} = E_{A} +\Omega_{A} +\Phi_{A}
\end{align}
where
\begin{align}
E_{A} =E_{A}{}^{M}\partial_{M}
,~~~
\Omega_{A} =  \tfrac12 \Omega_{A}{}^{bc} M_{b c}
,~~~
\Phi_{A}=  \Phi_{A}{}^{ij} J_{ij}
\end{align}
are the coframe, spin connection, and $SU(2)$ connection, respectively.
The generators of the superalgebra $\mathfrak{spin}(5,1)\oplus \mathfrak {sp}(1)\subset \mathfrak{osp}(6,2|1)$ are defined by their action on the spinors as
\begin{align}
\label{LorentzSpin}
[ M_{a b}, \mathcal D_{\gamma k} ] = -{ \tfrac12} ({\gamma}_{ab})_{\gamma}{}^{\delta} \mathcal D_{\delta k}
&~~\mathrm{and}~~
[ J_{ij}, \mathcal D_{\gamma k} ] = - \varepsilon_{k(i} \mathcal D_{\gamma j)}.
\end{align}
The graded commutation relations of the covariant derivatives define torsions, curvatures, and field strengths
\begin{align}
\label{TRF}
[ \mathcal D_{A}, \mathcal D_{B} \} &= T_{AB}{}^{C} \mathcal D_{C} + \tfrac12 R_{A B}{}^{c d} M_{cd} +F_{A B}{}^{ij} J_{ij}.
\end{align}
We will work with the supergeometry defined by the relations 
\begin{align}
\hspace{-1cm}
\label{E:DsDs}
\{ \mathcal D_{\alpha i}, \mathcal D_{\beta j} \}&= 
	2 i \varepsilon_{ij} (\gamma^{a})_{\alpha \beta}\mathcal D_{a}
	+2iC_{a \, ij} ({\gamma}^{a b c})_{\alpha \beta} M_{b c}
	+4i\varepsilon_{ij}N^{abc} (\gamma_{a})_{\alpha \beta} M_{bc}\cr
	&\hspace{3cm}
	-6i\varepsilon_{ij}C_{a}^{kl} (\gamma^{a})_{\alpha \beta} J_{kl} 
	-  \tfrac{8i}3 N^{a b c} ({\gamma}_{a b c})_{\alpha \beta}  J_{ij}\cr
%%%
[\mathcal D_{\gamma k}, \mathcal D_{a} ] &=
	C^{b l}_{k}({\gamma}_{ab})_{\gamma}{}^{\delta}  \mathcal D_{\delta l} 
	+N_{abc}({\gamma}^{bc})_{\gamma}{}^{\delta} \mathcal D_{\delta k}
	+\tfrac12R_{\gamma k a}{}^{bc}M_{bc}
\cr
	&+\left( (\gamma_a)_{\gamma \delta} \mathcal C^{\delta\,ij}_{k} 
		-6\delta^{i}_k \mathcal C_{a\,\gamma}{}^{j}
		+5\delta^{i}_k (\gamma_a)_{\gamma \delta}\left[ \mathcal C^{\delta j} -\tfrac13 \mathcal N^{\delta j}\right]\right) J_{ij}.
\end{align}
The curvature term of dimension $\tfrac32$ is an unilluminating function of the dimension-$\tfrac32$ torsion so we do not reproduce it here. The dimension-$\tfrac32$ torsion components $\mathcal C$ and $\mathcal N$ appear in the higher components of the $p$-forms. Their definitions are as the irreducible components 
\begin{align}
\mathcal D_{\gamma k } C_{a\, ij} &= \mathcal C_{a\, \gamma k\, ij} + (\gamma_a)_{\gamma \delta}\mathcal C^{\delta}_{ijk}
	+\varepsilon_{k(i}\mathcal C_{a\, \gamma j)} + \varepsilon_{k(i}(\gamma_a)_{\gamma \delta}\mathcal C^{\delta}_{j)}\cr
\mathcal D_{\gamma k} N_{\alpha \beta}&=\mathcal N_{\gamma k\, \alpha \beta}
	+\check{\mathcal N}_{\gamma k\, \alpha \beta}
\cr
\mathcal D_{\gamma k} N^{\alpha \beta}&= \mathcal N_{\gamma k}{}^{\alpha \beta} + \delta_\gamma^{(\alpha}\mathcal N^{\beta)}_k .
\end{align}
These components are constrained by the supergravity Bianchi identities to be \cite{Linch:2012zh, CesarThesis}
\begin{align}
\label{E:DCDN}
\begin{array}{ll}
\mathcal C_{a\, \gamma k\, ij}  = 0 
	&\mathcal N_{\gamma k\, \alpha \beta }=0\\
\mathcal C^\delta_{ijk}=- \tfrac 16 (\tilde \gamma^b)^{\delta \beta} \mathcal D_{\beta(k} C_{b\, ij)}
	&\check{\mathcal N}_{\gamma k \, \alpha \beta}= - \tfrac34 (\gamma^a)_{\gamma(\alpha} \mathcal C_{a\,\beta)k}\\
\mathcal C_{a\, \beta j}=  \tfrac23\Pi_{a\, \beta}^{c\, \gamma} \mathcal D^i_\gamma C_{a\, ij}
	& \mathcal N_{\gamma k}{}^{\alpha \beta} = \mathcal D_{\gamma k} N^{\alpha \beta} 
	- \tfrac25 \delta_\gamma^{(\alpha} \mathcal D_{\delta k}  N^{\beta)\delta}\\	
\mathcal C^{\gamma k}= - \tfrac19 \mathcal D_{\delta l} C^{\delta\gamma\, lk}
	&\mathcal N^{\alpha i} =  \tfrac 25 \mathcal D^i_\beta  N^{\beta\alpha} ,
\end{array}
\end{align}
where 
\begin{align}
\label{E:Projector}
	\Pi_{a\alpha}^{b\beta} = \delta_a^b\delta_\alpha^\beta + \tfrac16 (\gamma_a\tilde \gamma^b)_\alpha{}^\beta
\end{align}
is the projector onto the $\gamma$-traceless subspace of the spinor-vector representation: $\Pi_{a\alpha}^{b\beta} (\gamma_b)_{\beta \gamma} \equiv 0$ and $(\gamma^a)^{\gamma \alpha} \Pi_{a\alpha}^{b\beta} \equiv 0$.

Super-Weyl transformations preserving this geometry are generated by a real, unconstrained, scalar superfield $\sigma(z)$. The transformations that preserve the covariant derivative algebra act as
\begin{align}
\label{E:WeylTransformation}
\delta \mathcal D_{\alpha i} &= \tfrac12 \sigma \mathcal D_{\alpha i}
	-2(\mathcal D_{\beta i} \sigma)M_{\alpha}{}^{\beta} 
	+4(\mathcal D_{\alpha}{}^{j}\sigma) J_{ij} \\
\delta \mathcal D_{a} &= \sigma \mathcal D_{a} 
	-\tfrac i2 (\mathcal D^k\sigma)\tilde \gamma_{a}\, \mathcal D_k 
	-  (\mathcal D^{b}\sigma ) \, M_{ab}
	- \tfrac i8 (\mathcal D^i \tilde\gamma_{a} \mathcal D^j\sigma) \, J_{ij} 
\end{align}
on the covariant derivatives and as 
\begin{align}
\label{E:CWeylTransformation}
	\delta C_{a\, ij} &= \sigma C_{a\, ij} + \tfrac{i}{8}(\mathcal D_{(i} \tilde \gamma_{c} \mathcal D_{j)}\sigma)\\
\label{E:NWeylTransformation}
	\delta N_{abc} &=\sigma N_{abc} - \tfrac {i}{32}  (\mathcal D^k \tilde \gamma_{abc} \mathcal D_k \sigma)
\end{align}
on the dimension-1 torsions. 

In section \ref{S:Superforms} we introduce a commuting spinor $s^{\alpha i}$ that plays the role of $d\theta^{\alpha i}$ in the algebra of exterior superforms. The product of two of such basis elements decomposes into two parts
\begin{align}
s^{\alpha i} s^{\beta j} = 
	-\tfrac18\ \varepsilon^{ij} (\tilde \gamma_a)^{\alpha \beta} \gamma^a(s,s) 
	+\tfrac1{48} (\tilde \gamma^{abc})^{\alpha \beta}  \omega^{ij}_{abc}(s,s) 
\end{align}
where we have defined the vector and self-dual 3-form components
\begin{align}
\label{E:gammaomega}
	\gamma^a(s,s) := s^k \gamma^a s_k
	~~~\mathrm{and}~~~
	\omega^{ij}_{abc}(s,s) := s^{(i}\gamma_{abc} s^{j)} .
\end{align}
The vector $\gamma^a(s,s)$ is null as follows from the general identity 
\begin{align}
\label{E:FundamentalFierz}
(s^i \gamma^a s_i ) (s^j \gamma_a \xi_j )\equiv 0 
\end{align}
which holds for any chiral spinor $\xi$ (because $(\gamma^a)_{\alpha \beta} (\gamma_a)_{\gamma \delta} = 2 \varepsilon_{\alpha \beta \gamma \delta}$ whereas the isospin indices range only over 2 values). 
It is also orthogonal to the triplet of 3-forms $\gamma^a(s,s)\, \omega^{ij}_{abc}(s,s) \equiv 0$.
Many Fierz identities can be derived from these basic relations by polarizing on $s$ ({\it i.e.} replacing $s\to s+t+u$, expanding, and collecting like powers).

Projecting $s \mapsto \lambda \otimes v$ to the product of a commuting chiral spinor $\lambda^\alpha$ and isotwisor $v^i$
kills the vector part and isolates the self-dual 3-form part of the bilinear. A chiral spinor with the property 
\begin{align}
\label{E:PureSpinor}
\lambda^\alpha \lambda^\beta  = \tfrac 1{ 3! 2^3} (\tilde \gamma^{abc})^{\alpha \beta}  \lambda \gamma_{abc}\lambda
\end{align}
is called {\em pure} (see {\it e.g.} reference \cite{Berkovits:2004bw}) so we will refer to this projection as the pure spinor projection.

The constant, commuting spinor $s$ combines with the covariant derivative to define the odd derivation $\mathcal D_s = {s}^{\alpha i} \mathcal D_{\alpha i}$ which squares to
\begin{align}
\label{E:Ds^2}
\mathcal D_s^2 = 
i \mathcal D_{\gamma(s,s)} 
	+2i N_{{\gamma(s,s)}ab} M^{ab}
	-3iC_{\gamma(s,s)}^{ij} J_{ij}\cr
+  i \omega^{abc}_{ij}(s,s)
	\left( 	C^{ij}_c M_{ab}
	-\tfrac 43 N_{abc} J^{ij}
	\right).
\end{align}
In section \ref{S:Superforms}, we define the projected derivation $Q= \mathcal D_{\lambda \otimes v}$. Its square reduces to 
\begin{align}
\label{E:Q^2}
Q^2 = i (\lambda \gamma^{abc} \lambda) v^iv^j\left( C_{c\, ij} M_{ab}-\tfrac 43 N_{abc} J_{ij}\right) .
\end{align}
In section \ref{S:Conformal}, we introduce a complex of spaces on which this square vanishes, thereby promoting $Q$ to a differential.

%%%%%%%%%%%%%%%%%%%%%%%%%%%%%%%%%%%%%%%%%%%%%%%%%%%%%%%%%%%%%%%%%%
%%%%%%%%%%%%%%%%%%%%%%%%%%%%%%%%%%%%%%%%%%%%%%%%%%%%%%%%%%%%%%%%%%
\section{The non-abelian tensor hierarchy}
\label{S:NonAbTensorHierarchy}

In section \ref{S:TensorHierarchy}, we describe an application of the results on the structure of fundamental and composite $p$-forms to the (abelian part of the) non-abelian tensor hierarchy \cite{Samtleben:2011fj, Samtleben:2012mi}. 
In this appendix, we review very briefly the underlying cohomological structure of this hierarchy and propose a reformulation of it in terms of a generalized field strength satisfying a Maurer-Cartan equation.

The non-abelian tensor hierarchy is built on a collection of $p$-form gauge fields $(A^r, B^I, C_r, D_\alpha, E_\mu)$ where $p=1,2,3,4,5$, respectively. The representation indices take values in a collection of vector spaces that fit into a chain complex
\begin{align}
K_\bullet  = \dots \longrightarrow K_3
	\stackrel \partial \longrightarrow 
K_2 
	\stackrel \partial \longrightarrow 
K_1
	\stackrel \partial \longrightarrow 
K_0
\longrightarrow 0.
\end{align}
Here the representation space of a gauge $p$-form is denoted by $K_{p-1}$. The first few terms of the differential are denoted by $(\partial_1, \partial_2, \partial_3) = (\mathsf h, \mathsf g, \mathsf k)$. 

Introduce the de Rham complex 
\begin{align}
\Omega^\bullet  = 0 \longrightarrow \Omega^0
	\stackrel d \longrightarrow 
\Omega^1
	\stackrel d \longrightarrow 
\Omega^2
	\stackrel d \longrightarrow 
\Omega^3
\longrightarrow \dots
\end{align}
and consider the double complex $\Omega^\bullet \otimes K_\bullet$ with differential $D = d+\partial$. Let $\mathsf A$ denote an element of total degree 1 so that $\mathsf A$ represents the collection $(A^r, B^I, C_r, D_\alpha, E_\mu)$ of bi-degrees $(1,0), (2, -1), (3, -2), (4, -3), (5,-4)$, respectively. Then 
\begin{align}
\mathsf F : = D\mathsf A
\end{align}
is a form of total degree 2 representing $(F^r, H^I, G_r, K_\alpha, L_\mu)= (dA^r+ \mathsf h^r_I B^I, dB^I+ \mathsf g^{Ir}C_r, dC_r+ \mathsf k^\alpha_r D_\alpha, dD_\alpha+ \mathsf l^\mu_\alpha E_\mu, dE_\mu)$. Then 
\begin{align}
\label{E:DF}
D\mathsf F=0
	~~~\Leftrightarrow ~~~
\left\{
	\begin{array}{lcl}	
	dF^r &=&  \mathsf h^r_I H^I \\
	dH^I &=&  \mathsf g^{Ir}G_r \\
	dG_r &=&  \mathsf k^\alpha_r K_\alpha\\
	dK_\alpha &=&  \mathsf l^\mu_\alpha L_\mu \\
	dL_\mu&=& 0.
	\end{array}
\right.	 
\end{align}

Analogously to the algebra structure given to the de Rham complex by the wedge product, it turns out that $K_\bullet$ can be given an algebra structure by defining a collection of maps \cite{Samtleben:2011fj}
\begin{align}
* : K_p \times K_q \to K_{p+q+1} .
\end{align}
When we wish to distinguish them, will denote the non-vanishing restrictions of the product by $(\mathsf f^t_{rs}, \mathsf d^I_{rs}, \mathsf b_{Irs}, \mathsf c_{\alpha IJ}, \mathsf c^{\prime s}_{\alpha r})$. Together with the tensors defining the differential on $K_\bullet$, these satisfy a list of identities shown in reference \cite{Palmer:2013pka} to make the $*$-product associative $(a*b) *c = a* (b*c)$ and the differential $\partial$ a derivation $\partial (a*b) = (\partial a) *b + a*(\partial b)$ of the resulting algebra $\forall a,b,c\in K_\bullet$. Here, we will extend this product to the double complex by wedge, that is, we interpret $*$ on the double complex to mean $*$ on $K_\bullet$ together with $\wedge$ on $\Omega^\bullet$. 

We are now in a position to use the $*$-product to deform the closure condition (\ref{E:DF}) on $\mathsf F$. To do this, one should first extend the differential on the double complex to a connection $\nabla = D + A*$. With this, one can define the non-abelian field strength $\mathsf F := \nabla * \nabla$. Finally, one writes $\nabla \mathsf F + \mathsf F* \mathsf F =0$.

The non-abelian tensor hierarchy {\em appears} to have a non-linear abelian limit obtained by setting $\mathsf f \to 0$ after deforming the complex with the $*$-product.\footnote{\label{F:Abelianization}The na\"ive limit appears to be non-trivial but we have not completed the analysis required to show the existence of non-trivial solutions to all the conditions this limit affects. Irrespective of this, the consistency of the resulting, perhaps formal, structure is a prerequisite to the extension to the full non-abelian hierarchy. The latter has been explicitly checked to have non-trivial solutions \cite{Samtleben:2011fj}. We thank Robert Wimmer for emphasizing these important points to us.} %end footnote
For simplicity, we will work in this limit. Then the deformed version of the closure condition (\ref{E:DF}) may be postulated to be the Maurer-Cartan equation
\begin{align}
D\mathsf F + \mathsf F*\mathsf F = 0 .
\end{align}
This equation expands out to 
\begin{align}
	dF^r &=  \mathsf h^r_I H^I \\
	dH^I &=  \mathsf g^{Ir}G_r  + \mathsf  d^I_{rs} F^r\wedge F^s\\
	dG_r &=  \mathsf k^\alpha_r K_\alpha  +  \mathsf b_{Irs} F^s\wedge H^I  \\
	dK_\alpha &=  \mathsf l^\mu_\alpha L_\mu +  \mathsf c_{\alpha IJ} H^I\wedge H^J +  \mathsf c^{\prime s}_{\alpha r} F^r \wedge G_s\\
	dL_\mu&= 0.
\end{align}
In section \ref{S:TensorHierarchy} we connect this construction to the sourced and composite $p$-form complices of sections \ref{S:deRham} and \ref{S:Composite}.

%%%%%%%%%%%%%%%%%%%%%%%%%%%%%%%%%%%%%%%%%%%%%%%%%%%%%%%%%%%%%%%%%%
%%%%%%%%%%%%%%%%%%%%%%%%%%%%%%%%%%%%%%%%%%%%%%%%%%%%%%%%%%%%%%%%%%
%%%%%%%%%%%%%%%%%%%%%%%%%%%%%%%%%%%%%%%%%%%%%%%%%%%%%%%%%%%%%%%%%%
%%%%%%%%%%%%%%%%%%%%%%%%%%%%%%%%%%%%%%%%%%%%%%%%%%%%%%%%%%%%%%%%%%

%%%%%%%%%%%%%%%%%%%%%%%%%%%%%%%%%%%%%%%%%%%%%%%%%%%%%%%%%%%%%%%%%%
%%%%%%%%%%%%%%%%%%%%%%%%%%%%%%%%%%%%%%%%%%%%%%%%%%%%%%%%%%%%%%%%%%

\end{document}